\documentclass[12pt]{article}
\usepackage{graphicx}
\usepackage{amssymb}
\usepackage{amsmath}
\usepackage{pgfplots}
\newtheorem{thm}{Theorem}[section]

\newtheorem{lem}[thm]{Lemma}

\newtheorem{pro}{Property}[section]

\newcommand{\A}{\mathcal{A}}
\numberwithin{equation}{section}
\DeclareMathOperator{\csch}{csch}
\DeclareMathOperator{\sech}{sech}
\DeclareMathOperator\Res{Res}
\DeclareMathOperator\sign{sign}
\DeclareMathOperator\diag{diag}
\DeclareMathOperator\col{col}

\newtheorem{Pa}{Paper}[section]

\newtheorem{La}[Pa]{{\bf Lemma}}
\newtheorem{Cy}[Pa]{{\bf Corollary}}
\newtheorem{Rk}[Pa]{{\bf Remark}}
\newtheorem{Pb}[Pa]{{\bf Problem}}
\newtheorem{Ee}[Pa]{{\bf Example}}
\newtheorem{Dn}[Pa]{{\bf Definition}}

\newtheorem{An}[Pa]{{\bf Assertion}}

\pgfplotsset{compat = newest}
\begin{document}

\title{On the dynamics of the singularities of the solutions of some non-linear integrable differential equations} 
\date{March 24, 2017}
\author{Igor Tydniouk}
\maketitle
\begin{center}\emph{Stevens Institute of Technology,\\
1 Castle Point Terrace, Hoboken, NJ 07030, USA}\\
 E-mail:itydniou@stevens.edu
\end{center}
\begin{center}\textbf{Abstract}\end{center}
This paper concerns with some of the results related to the singular solutions of certain types of non-linear integrable differential equations (NIDE) and behavior of the singularities of those equations. The approach heavily relies on the \emph{Method of Operator Identities} \cite{lev1} which proved to be a powerful tool in different areas such as interpolation problems, spectral analysis, inverse spectral problems, dynamic systems, non-linear equations. We formulate and solve a number of problems (direct and inverse) related to the singular solutions of sinh-Gordon, non-linear Schr\"{o}dinger and modified Korteweg - de Vries equations. Dynamics of the singularities of these solutions suggests that they can be interpreted in terms of particles interacting through the fields surrounding them. We derive differential equations describing the dynamics of the singularities and solve some of the related problems. The developed methodologies are illustrated by numerous examples. 
\\
\section{Introduction}
\emph{Method of Operator Identities} \cite{lev1} plays an important role in different areas of both pure and applied mathematics. This method appeared to be a 
universal tool for solving the interpolation, spectral analysis problems, investigation of dynamic systems and nonlinear integrable equations. Solutions of many problems that became already classical are much simpler and more transparent under the prism of \emph{Method of Operator Identities} and it reveals the striking similarities between very different at the first glance fields of research. \\
In this paper we apply \emph{Method of Operator Identities} to the investigation of the properties of the singular solutions of some non-linear integrable equations obtained by solving the inverse spectral problem for the associated self-adjoint \emph{canonical system} of differential equations. In particular, we consider the following non-linear equations
\begin{equation} \label{eq:SHG-0}
\frac {\partial ^2 \phi (x, t)}{\partial x \partial t} = 4 \sinh \phi (x, t) \quad \text{sinh-Gordon equation (SHG) };
\end{equation}
\begin{equation} \label{eq:MKDV-0}
\begin{split}
\frac {\partial \psi (x, t)}{\partial t} = - \frac{1}{4} \frac {\partial ^3 \psi (x, t)}{\partial x ^3} +
\frac{3}{2} | \psi (x, t) |^2 \frac {\partial \psi (x, t)}{\partial x}
\\
\text{modified Korteweg - de Vries equation (MKdV)};
\end{split}
\end{equation}

\begin{equation} \label{eq:NSE-0}
\begin{split}
\frac {\partial \rho (x, t)}{\partial t} = \frac{\imath}{2} \left [\frac {\partial ^2 \rho (x, t)}{\partial x ^2} -
2 | \rho (x, t) |^2 \rho (x, t) \right ]
\\
\text{non-linear Schr\" {o}dinger equation (NSE)}.
\end{split}
\end{equation}

Some of the results concerning these equations were obtained previously by different methods but for the completeness of the picture and to show the universality and power of the \emph{Method of Operator Identities} we present the solutions and proofs here. The main subject of investigation is the study of the properties of the singular solutions and the behavior of the singularities of those solutions. Initially the idea of investigation of singular solutions was suggested in \cite{pogreb1, pogreb2} where the "gluing" procedure was applied to the inverse scattering problem as a method of analysis. Method of \emph{Inverse Spectral Problem} powered by \emph{Method of Operator Identities} proved to be more efficient in these investigations and allowed to perform more general and more detailed analysis of the considered solutions. The properties of the singular solutions (already discussed in \cite{pogreb1, pogreb2}) point out that on global scale they behave very similar to the classical soliton solutions: asymptotically $N$-wave solution is represented as $N$ independent elementary waves; after the interaction elementary waves preserve their shapes and the only change they experience is
 the phase shift; during the interaction elementary waves exchange their energies. Singular solutions admit interpretation in terms of particles interacting through the fields surrounding them. As opposed to the soliton solutions, presence of the singularities allows to derive dynamical equations and investigate in much more details the region of close interaction between singular waves/particles.\\
The plan of the paper is as follows. \textbf{Section 2} is auxiliary. There we introduce a class of structured matrices (paired Cauchy matrices) related to the equations \eqref{eq:SHG-0}-\eqref{eq:NSE-0} and using \emph{Method of Operator Identities} (matrix version) we investigate the invertibility of these matrices and calculate the transfer matrix function of the corresponding dynamic system. Studying the properties of the dynamic system led us to the investigation of some related rational direct and inverse interpolation problems. Obtained results are applied in further sections to study the properties of the singularities of non-linear equations. At the same time, results of the \textbf{Section 2} are of independent interest in the field of structured matrices and related interpolation problems. In particular, we investigate the following interpolation problem\\
\textbf{IP Problem.} \emph{Given the sets of numbers
$$
\mu = \{\mu_1, \mu_2, \ldots, \mu_n\}, \; \nu = \{\nu_1, \nu_2, \ldots, \nu_n\}, \; \xi =\{\xi_1, \xi_2, \ldots, \xi_n\},
$$
 find $2 \times 2$ matrix polynomial $X(\lambda) = \left \{ X_{i j}(\lambda) \right \}_{i, j = 1}^2$  satisfying the relations} 
\begin{equation} \label{eq:IPP-0}
X(\xi_j)
\begin{bmatrix}
  \nu_j  \\
  \mu_j  
 \end{bmatrix} = 0, \quad 1 \leq j \leq n.
\end{equation}
This and similar interpolation problems were studied by the number of the authors (see for example \cite{heinig2} and \cite{gohberg1} - \cite{saed}, \cite{ball}). The use of the \emph{Method of Operator Identities} reveals some interesting connecting links among different areas of analysis such as dynamic systems, structured matrices and non-linear differential equations.\\ 
In \textbf{Section 3} (\textbf{Subsection 3.1}) we consider explicit singular solutions of non-linear integrable equations. The procedure relies on the operator version of the \emph{Method of Operator Identities}. It is shown that those solutions can be represented in terms of determinants of the paired Cauchy and paired Vandermonde matrices (\textbf{Theorems 3.3} and \textbf{3.4}).

In \textbf{Subsection 3.2} we study the properties of the singular solutions. Using results of \textbf{Section 2} we obtain an efficient parametrization of the zeros of those determinants and investigate the connection between transfer
 matrix function of the corresponding dynamic system and singular solutions of non-linear equations (\textbf{Theorem 3.5}). In this way we formulate and solve an inverse problem of singular solutions: given some information about the solution, restore the full system (\textbf{Theorems 3.6, 3.7} and \textbf{3.8}). Developed methodologies are illustrated by simple examples.

 \textbf{Subsection 3.3} is dedicated to the investigation of the dynamics of the singularities given by the parametrizations obtained in \textbf{Subsection 3.2}. It is shown that the dynamics of the singularities is described by completely integrable Hamiltonian system and \emph{action-angle} variables for this system are found (\textbf{Theorem 3.11}). We also derive a system of non-linear differential equations describing the dynamics of the parameters and study the properties of the system for some special simple cases (2-wave interaction). Numerous examples showing different aspects of the solutions are presented. For the case of two-wave interaction we formulate and solve an inverse problem (\textbf{problem 3.35}, \textbf{Assertion 3.36}). In general ($N$-wave interaction), dynamics of the singularities is quite complicated and cannot be integrated in closed form. In \textbf{Appendix} we present some of the examples of singularities behavior obtained by numerical analysis and give an interpretation in terms of particles.

\section*{Acknowledgements}
I am deeply grateful to Dr. A.L. Sakhnovich for carefully reading this paper and correcting numerous typos and mistakes. His ideas and insights had a crucial influence on my way of thinking.

\section{Dynamic systems, operator identity and associated interpolation problems}
Consider matrix $S$ of the form
\begin{equation} \label{eq:S}
        S=\left\{ \frac{a_i b_j + c_i d_j}{g_i - h_j} \right\} ^N _{i,j=1},
\end{equation}
where
 $ a=\{ a_i \} ^N _1 , \; b=\{ b_i \} ^N _1 , \; c=\{ c_i \} ^N _1 , \;
d=\{ d_i \} ^N _1 , \; g=\{ g_i \} ^N _1 , \; h=\{ h_i \} ^N _1
 $
 - are the sets of complex numbers such that
 $
 g_i \neq h_j \; \left( 1\leq i,
j \leq N \right) ; \; g_i \neq g_j, \; h_i \neq h_j, \; i \neq j
 $.\\
In the special case when
$
a_i b_j \: + \: c_i d_j = 1 \left( 1 \leq i, j
\leq N \right)
$
matrix $S$ is a pure Cauchy matrix. Matrices of the type~\eqref{eq:S} represent a special case of generalized Cauchy matrices in the sense of \cite{heinig1}. They were studied by the number of the authors (see for example \cite{heinig2} and \cite{gohberg1} - \cite{saed}). Numerous interpolation problems connected to the matrices of this class were investigated in \cite{ball}. Results of this section slightly generalize the ones obtained in \cite{heinig3}. Our approach is
 based on matrix identity
\begin{equation} \label{eq:identity}
        AS-SB=\Pi _1 \Pi ^T _2,
\end{equation}
where
$
A = \diag\{g_1 , g_2 , \ldots g_N \} , \; B=\diag\{h_1 , h_2 ,
\ldots h_N \},
$
\\

\begin{center}
$
        \Pi_1 = \begin{bmatrix}
                      a_1 & c_1 \\
                      a_2 & c_2 \\
											\cdots & \cdots\\
											a_N & c_N \\
                    \end{bmatrix},\;
        \Pi_2 = \begin{bmatrix}
                      b_1 & d_1 \\
                      b_2 & d_2 \\
											\cdots & \cdots\\
											b_N & d_N \\
                    \end{bmatrix};
$
\end{center}
and the symbol $M^T$ denotes transposition of the matrix $M$.
This is a matrix version of operator identity thoroughly investigated and used in \cite{lev1}, \cite{sasa1} and a number of papers (see for example \cite{lev2} - \cite{sasa4}). In this section we review the results related to the rational interpolation problems and invertibility of the matrices of type~\eqref{eq:S} which play an important role in further considerations concerning singular solutions of NIDE.    

Let's introduce $2 \times 2$ matrix-function $W_A \left( \lambda
\right)$ by the equality
\begin{equation} \label{eq:TMFA}
       W_A \left( \lambda \right) = I_2 -\Pi ^T _2 S^{-1} \left(
       A-\lambda I_N \right) ^{-1} \Pi_1,
\end{equation}
where $I_k$ - is the $k \times k$ identity matrix. Note that $W_A \left(
\lambda \right)$ is transfer matrix-function of the dynamic system
\begin{equation} \label{eq:DynS}
       \frac{dx}{dt} = Ax+\Pi_1 u, \; y=\Pi ^T _2 S^{-1} x + u,
\end{equation}
where $u=\{u_i \left( t \right) \} ^2 _1$ - input, $y=\{y_i \left( t
\right) \} ^2 _1$ - output, and $x=\{x_i \left( t \right) \} ^N _1$
- is the inner state of the system. Matrix-function $W_B \left(
\lambda \right) = W^{-1} _A \left( \lambda \right)$ that can be
represented in the form \cite{lev1}
\begin{equation}\label{eq:TMFB}
       W_B \left( \lambda \right) = I_2 +\Pi ^T _2 \left(
       B-\lambda I_N \right) ^{-1} S^{-1} \Pi_1
\end{equation}
also plays an important role in the following studies. As one can
see from~\eqref{eq:TMFA},~\eqref{eq:TMFB} the existence of $W_A \left( \lambda \right)$ and $W_B \left( \lambda \right)$ depends on the invertibility 
of the matrix $S$. Let's define the ordered sets
\begin{center}
$
        \mu = \{\mu_i  \} = \left[a_1 , a_2 , \ldots a_N , d_1 , d_2 , \ldots d_N,
        \right],
$
\end{center}
\begin{center}
$
        \nu = \{\nu_i  \} = \left[-c_1 , -c_2 , \ldots -c_N , b_1 , b_2 , \ldots b_N
        \right],
$
\end{center}
\begin{center}
$
        \xi = \{\xi_i  \} = \left[g_1 , g_2 , \ldots g_N , h_1 , h_2 , \ldots h_N \right].
$
\end{center}
The criteria of regularity of the matrix $S$ is given by the following theorem\\
\begin{thm}
Let a matrix $S$ have the form~\eqref{eq:S} and assume that $\mu_{i_k} = 0$ and $\nu_{i_k} = 0$ for some sets $i_1, i_2, \cdots, i_p$ and $j_1, j_2, \cdots, j_r$ of natural numbers (less or equal $2N$) such that $0 \leq p \leq 2N, \; 0 \leq r \leq 2N$ and $i_k \neq j_m$ for all $1 \leq k \leq p$ and $1 \leq m \leq r$. Then the relations 
\begin{equation} \label{eq:param26}
    \mu_n Q_1 \left( \xi_n \right) + \nu_n Q_2 \left( \xi_n \right) = 0, \; n=1, 2, \ldots , 2N
\end{equation}
with some polynomials $Q_1$ and $Q_2$ of the form
\begin{center}
$ Q_1 \left( \lambda \right) = \tilde{Q_1} \left( \lambda \right)
\prod^{r}_{m=1} \left( \lambda - \xi_{j_m} \right)$,
\end{center}
\begin{center}
 $Q_2 \left( \lambda \right) = \tilde{Q_2} \left( \lambda \right)
\prod^{p}_{k=1} \left( \lambda - \xi_{i_k} \right)$,
\end{center}
where $\tilde{Q_l} \left( \lambda \right) \left( l=1,2 \right)$ are arbitrary polynomials such that
\begin{center}
$ \deg\{\tilde{Q_1} \left( \lambda \right)\} \leq N-1-r, \; \; \deg \{
\tilde{Q_2} \left( \lambda \right)\} \leq N-1-p$,
\end{center}
are necessary and sufficient for the matrix $S$ to be singular i.e. $\det S = 0.$
\end{thm}

\begin{Rk}
The proof of the \textbf{Theorem 2.1} can be easily obtained from the results of \cite{heinig3}. We give it here for the completeness of the considerations. 
\end{Rk}
\emph{Proof.} Assume for now that $\mu_i \neq 0, \nu_i \neq 0; i = 1, 2, \dots , 2N$. Condition $\det S = 0$ is equivalent to the existence of the non-trivial solution $x = \{x_i\}_1 ^N$ of the system of equations $S x = 0$ or
\begin{equation} \label{eq:sys27}
\sum_{j = 1}^{N} {\frac {a_i b_j - c_i d_j}{g_i - h_j}x_j} = 0, i = 1, 2, \dots , N.   
\end{equation}
System~\eqref{eq:sys28} can be rewritten as
\begin{equation} \label{eq:sys28}
a_i \sum_{j = 1}^{N} {\frac {b_j x_j}{g_i - h_j}} = y_i, \; 1 \leq i \leq N,   
\end{equation}
\begin{equation} \label{eq:sys29}
c_i \sum_{j = 1}^{N} {\frac {d_j x_j}{g_i - h_j}} = y_i, \; 1 \leq i \leq N,   
\end{equation}
where $y = \{y_i\}_1^N$ - non-trivial vector. Consider functions
\begin{equation} \label{eq:sys210}
\begin{aligned}
G_1(\lambda) = \sum_{j = 1}^{N} {\frac {b_j x_j}{\lambda - h_j}} = \frac {\sum_{j = 1}^{N} {b_j x_j \prod_{k \neq j}{(\lambda - h_k)}}} {\prod_{i = 1}^N {(\lambda - h_i)}} = \frac {f_1(\lambda)}{H(\lambda)}, \\  
G_2(\lambda) = \sum_{j = 1}^{N} {\frac {d_j x_j}{\lambda - h_j}} = \frac {\sum_{j = 1}^{N} {d_j x_j \prod_{k \neq j}{(\lambda - h_k)}}} {\prod_{i = 1}^N {(\lambda - h_i)}} = \frac {f_2(\lambda)}{H(\lambda)}.  
\end{aligned}
\end{equation}
From~\eqref{eq:sys28},~\eqref{eq:sys29} it follows that $a_i G_1(g_i) = c_i G_2(g_i) = y_i; \; 1 \leq i \leq N$. Substituting in~\eqref{eq:sys210} $\lambda = h_j$ we obtain
$$
f_1(h_j)=b_j x_j H'(h_j),
$$

$$
f_2(h_j)=d_j x_j H'(h_j),
$$
or
\begin{equation} \label{eq:sys211}
f_1(h_j) d_j = f_2(h_j) b_j, \; 1 \leq j \leq N.
\end{equation}
It follows from~\eqref{eq:sys210} that $deg f_1(\lambda) \leq N - 1; \; deg f_2(\lambda) \leq N - 1$. On the other hand, using~\eqref{eq:sys210},  expressions~\eqref{eq:sys28},~\eqref{eq:sys29} can be represented as 
$$
y_i = a_i \frac {f_1 (g_i)}{H(g_i)}, \; 1 \leq i \leq N,
$$
$$
y_i = c_i \frac {f_2 (g_i)}{H(g_i)}, \; 1 \leq i \leq N,
$$
or
\begin{equation} \label{eq:sys212}
a_i f_1 (g_i) = c_i f_2 (g_i), \; 1 \leq i \leq N.
\end{equation}
Formulas~\eqref{eq:sys211},~\eqref{eq:sys212} prove the necessity of the conditions of the theorem in the case $\mu_i \neq 0, \; \nu_i \neq 0; \; i = 1, 2, \dots , 2N$. Reverse considerations give the sufficiency. Let now $\mu_{i_k} = 0, \; k = 1, 2, \dots, p$ for some multi-index $i_1 , i_2, \dots i_p, \; 0 \leq p \leq 2N$. Then from~\eqref{eq:sys211},~\eqref{eq:sys212} it follows that $f_2 (\xi_{i_k}) = 0, \; k = 1, 2, \dots, p$. In other words, $f_2 (\lambda) = \tilde{f_2} (\lambda) \prod_{k = 1}^p {(\lambda - \xi_{i_k})}$ and $\tilde{f_2}(\lambda)$ is the polynomial such that $deg \tilde{f_2}(\lambda) \leq N - 1 - p$. If $\nu_{j_m} = 0, \; m = 1, 2, \dots, r$ for some multi-index $j_1 , j_2, \dots j_r, \; \; 0 \leq r \leq 2N$, then $f_1 (\xi_{j_m}) = 0, \; m = 1, 2, \dots, r$ and $f_1 (\lambda) = \tilde{f_1} (\lambda) \prod_{m = 1}^r {(\lambda - \xi_{j_m})}$ where $\tilde{f_1}(\lambda)$ is the polynomial such that $deg \tilde{f_1}(\lambda) \leq N - 1 - r$. It's easy to see that equalities $i_k = j_m $ for any of the pairs $(k, m); \; k = 1, 2, \dots, p; \; m = 1, 2, \dots, r$ are impossible because in these cases the determinant of the matrix $S$ equals zero. 
$\Box$
\begin{Rk}
Equations~\eqref{eq:param26} parametrize the
equality $\det S = 0$ by means of the coefficients of the polynomials
$ \tilde{Q_1} \left( \lambda \right)$ and $\tilde{Q_2} \left( \lambda \right).$
\end{Rk}
The parametrization is understood in the following sense. Let $\left\{q_{i}^{(1)}\right\}_{i = 1}^{N - r}$ and $\left\{q_{i}^{(2)}\right\}_{i = 1}^{N - p}$ represent the coefficients of the polynomials $\tilde{Q_1}(\lambda)$ and $\tilde{Q_2}(\lambda)$ respectively then considering $\det S$ as a function 
$$
F\left(q_{1}^{(1)}, q_{2}^{(1)}, \ldots, q_{N - r}^{(1)}, q_{1}^{(2)}, q_{2}^{(2)}, \ldots, q_{N - p}^{(2)}\right)
$$ of the parameters $\left\{q_{i}^{(1)}\right\}$ and $\left\{q_{i}^{(2)}\right\}$, we have

$$
F\left(q_{1}^{(1)}, q_{2}^{(1)}, \ldots, q_{N - r}^{(1)}, q_{1}^{(2)}, q_{2}^{(2)}, \ldots, q_{N - p}^{(2)}\right) = 0,
$$
which can also be considered as equation of the surface in $(2 N - 2 - r - p)$ - dimensional space.\\
Let the sets $ \mu , \nu , \xi $ be such that $ \det S \neq 0$. Then
from~\eqref{eq:TMFA} it follows that
\begin{equation} \label{eq:sys213}
    W_A \left( \lambda \right) = \left (\prod^{N}_{i=1} \left( g_i -
    \lambda \right) ^{-1}\right ) \{ D_{jk} \left( \lambda \right)
    \}^2_{j,k=1},
\end{equation}
where $ D_{jk} \left( \lambda \right) \; \left( j, k = 1, 2 \right)$
- are polynomials such that
\begin{center}
    $\deg \{ D_{11} \left( \lambda \right) \} \leq N; \; \deg \{ D_{22} \left( \lambda \right) \} \leq
    N$,
\end{center}
\begin{center}
    $\deg \{ D_{21} \left( \lambda \right) \} \leq N - 1; \; \deg \{ D_{12} \left( \lambda \right) \} \leq
    N - 1$.
\end{center}

We now formulate and solve related interpolation problems. Note, that the similar problems were considered in \cite{ball}. The proofs become much simpler
 and more transparent if one uses identity~\eqref{eq:identity} and general expression for the transfer matrix-function $W_A(\lambda)$. Relations~\eqref{eq:identity} and~\eqref{eq:TMFA} allow a unified approach to the problems from different areas, i.e. dynamic systems, interpolation, spectral problems, non-linear differential equations, as we'll see in the following sections.\\
Let's introduce the projectors $P_k \; (1 \leq k \leq N)$ as $N \times N$ matrices defined by
$$
P_k = \left\{p_{ij}\right\}_{i, j = 1}^N: \; p_{ij} = 0 \; \text{when} \; i \neq j \; \text{and} \; p_{ij} = 1 \; \text{when} \; i = j = k.
$$
Then
$$
\Pi_2^T P_k =
\begin{bmatrix}
  0 & \cdots & 0 & b_k & 0 & \cdots & 0  \\
  0 & \cdots & 0 & d_k & 0 & \cdots & 0  
 \end{bmatrix}, \; 
1 \leq k \leq N.
$$
Multiplying from the right both sides of~\eqref{eq:TMFA} by $\Pi_2^T P_k$
we get
\begin{equation} \label{eq:sys214}
    W_A ( \lambda ) \Pi_2^T P_k = \left [\Pi_2^T - \Pi_2^T S^{-1} (A - \lambda E_N)^{-1} \Pi_1 \Pi_2^T \right]P_k, \; 1 \leq k \leq N.
\end{equation}
From~\eqref{eq:identity} it follows that 
\begin{equation} \label{eq:sys215}
    \Pi_1 \Pi_2^T = (A - \lambda E_N) S - S (B - \lambda E_N).
\end{equation}
Substituting~\eqref{eq:sys215} in~\eqref{eq:sys214} and passing to the limit $\lambda \to h_k$ results in 
\begin{equation} \label{eq:sys216}
    W_A \Pi_2^T P_k = 0, \; 1 \leq k \leq N.
\end{equation}
Taking into account~\eqref{eq:sys213}, equalities~\eqref{eq:sys216} can be written as
\begin{equation} \label{eq:sys217}
		b_k D_{i1} (h_k) + d_k D_{i2} (h_k) = 0; \; i = 1, 2; \; 1 \leq k \leq N.
\end{equation}
From~\eqref{eq:sys213} and relation $W_B^{-1} (\lambda) = W_A (\lambda)$ follows the representation 
\begin{equation} \label{eq:sys218}
		W_B ( \lambda ) = \frac{\prod_{i = 1}^N {(g_i - \lambda)}}{\det W_A(\lambda)} 
\begin{bmatrix}
  D_{22} (\lambda) & -D_{12} (\lambda)  \\
  -D_{21} (\lambda) & D_{11} (\lambda)  
 \end{bmatrix}.
\end{equation}
Now, multiplying from the left both sides of~\eqref{eq:TMFB} by $P_k^T \Pi_1$ and passing to the limit $\lambda \to g_k$
we obtain
\begin{equation} \label{eq:sys219}
P_k^T \Pi_1 W_B(g_k) = 0, \; 1 \leq k \leq N.
\end{equation}
Taking into account~\eqref{eq:sys218}, equalities~\eqref{eq:sys219} become
\begin{equation} \label{eq:sys220}
		a_k D_{i2}(g_k) - c_k D_{i1} (g_k) = 0; \; i = 1, 2; \; 1 \leq k \leq N.
\end{equation}

Expressions~\eqref{eq:sys217},~\eqref{eq:sys220} can be re-written in the form
\begin{equation} \label{eq:sys221}
\begin{bmatrix}
  D_{11} (\xi_k) & D_{12} (\xi_k)  \\
  D_{21} (\xi_k) & D_{22} (\xi_k)  
 \end{bmatrix}
\begin{bmatrix}
  \nu_k  \\
  \mu_k  
 \end{bmatrix} = 0, \; 
 1 \leq k \leq 2N.
\end{equation}
Equalities~\eqref{eq:sys221} can be reformulated in terms of the interpolation problem:\\\\
\textbf{IP Problem.} \emph{Given the sets of numbers $\mu, \nu, \xi$, find $2 \times 2$ matrix polynomial $X(\lambda) = \left \{ X_{i j}(\lambda) \right \}_{i, j = 1}^2$  satisfying the relations} 
\begin{equation} \label{eq:sys222}
X(\xi_j)
\begin{bmatrix}
  \nu_j  \\
  \mu_j  
 \end{bmatrix} = 0, \;
 1 \leq j \leq 2N.
\end{equation}
Let's note that this \textbf{IP Problem} has infinitely many solutions. Indeed, for any given vector polynomials $X_{k 1}(\lambda)$ (or $X_{k 2}(\lambda)$), $k = 1,2$ using~\eqref{eq:sys217} or~\eqref{eq:sys220}, Lagrange-Sylvester formulas give the way to recover corresponding polynomials $X_{k 2}(\lambda)$ (or $X_{k 1}(\lambda)$), $k = 1,2$.
From the set of the solutions of \textbf{IP Problem} we choose the one for which 
\begin{equation} \label{eq:sys223}
\deg X_{1 1}(\lambda) = \deg X_{2 2}(\lambda) = N; \; \deg X_{1 2}(\lambda) \leq N - 1; \; \deg X_{2 1}(\lambda) \leq N - 1.
\end{equation}
In this case the solution $\left \{ X_{i j}(\lambda) \right \}_{i, j = 1}^2$ is called the \emph{basis solution} of the \textbf{IP Problem} and $N$ is called the \emph{degree} ($\deg$) of the solution. The \emph{basis solution} $\left \{ X_{i j}(\lambda) \right \}_{i, j = 1}^2$ is called \emph{normalized basis solution} if the coefficients of the highest degree of the polynomials $X_{1 1}$ and $X_{2 2}$ are equal to 1.

The following considerations are devoted to the construction of the \emph{basis solution} of the \textbf{IP Problem}.\\ 
With the notations
\begin{center}
    $V_k \left( \eta,\zeta\right)=\begin{bmatrix}
                                    \eta_1 & \eta_2 & \cdots & \eta_{2N} \\
                                    \zeta_1 \eta_1 & \zeta_2 \eta_2 & \cdots & \zeta_{2N} \eta_{2N} \\
                                    \cdots & \cdots & \cdots & \cdots \\
                                    \zeta^k_1 \eta_1& \zeta^k_2 \eta_2 & \cdots & \zeta^k_{2N} \eta_{2N} \\
                                  \end{bmatrix};
 $
\end{center}
\begin{center}
    $   \Lambda_k = \col[1,\lambda, \cdots, \lambda^k]; \; \; V = \begin{bmatrix}
                                    V_{N - 1} \left( \mu , \xi \right) \\
                                    V_{N - 1} \left( \nu , \xi \right) \\
                                  \end{bmatrix}; \; \Delta = \det V$
\end{center}
we prove the following statements.
\begin{lem}
The matrix $V$ is non-singular if and only if the matrix $S$ is non-singular.
\end{lem}

\emph{Proof.} We rewrite matrix $S$ in terms of the sets $\mu$, $\nu$ and $\xi$ 
\begin{equation} \label{eq:sys2231}
        S=\left\{ \frac{\mu_i \nu_{N + j} - \nu_i \mu_{N + j}}{\xi_i - \xi_{N + j}} \right\} ^N _{i,j=1}
\end{equation}
and consider two related matrices
\begin{equation} \label{eq:sys2232}
        S_1=\left\{ \frac{\mu_i \nu_{N + j}}{\xi_i - \xi_{N + j}}\right\} ^N _{i,j=1}, \: 
				S_2=\left\{ \frac{ - \nu_i \mu_{N + j}}{\xi_i - \xi_{N + j}}\right\} ^N _{i,j=1}.
\end{equation}
These matrices can be represented as 
\begin{equation} \label{eq:sys2233}
        S_1 = M_1 S_0 N_2; \: S_2 = M_2 S_0 N_1,
\end{equation}
where $M_1$, $M_2$, $N_1$, $N_2$ are diagonal matrices
$$
M_1 = \diag \left [ \mu_1, \mu_2, \ldots, \mu_N \right ], \; M_2 = \diag \left [ \mu_{N + 1}, \mu_{N + 2}, \ldots, \mu_{2N} \right ];
$$
$$
N_1 = \diag \left [ \nu_1, \nu_2, \ldots, \nu_N \right ], \; N_2 = \diag \left [ \nu_{N + 1}, \nu_{N + 2}, \ldots, \nu_{2N} \right ];
$$
and $S_0$ is a Cauchy matrix
$$
S_0 = \left \{ \frac{1}{\xi_i - \xi_{N + j}}\right \}_{i, j = 1}^N.
$$
The determinants of the matrices $S_1$ and $S_2$ can be easily calculated as 
$$
\det S_1 =  \prod\limits_{i = 1}^N{\left (\mu_i \nu_{N + i} \right)} \det S_0;
$$
\begin{equation} \label{eq:sys2234}
\end{equation}
$$
\det S_2 = (-1)^N \prod\limits_{i = 1}^N{\left (\nu_i \mu_{N + i} \right)} \det S_0,
$$
where 
\begin{equation} \label{eq:sys2235}
\det S_0 = \frac{\prod\limits_{1 \leq i < j \leq N} {(\xi_i - \xi_j)} \prod\limits_{N + 1 \leq i < j \leq 2 N} {(\xi_i - \xi_j)}}{\prod\limits_{\substack{1 \leq i \leq N \\ N + 1 \leq j \leq 2 N}} {(\xi_i - \xi_j)}}.
\end{equation}
Let $\tau = \left \{ \tau_i \right \}_{i = 1}^N$ be an $N$-tuple of integers such that
\begin{enumerate}
\item $1 \leq \tau_i \leq 2 N;$
\item $\tau_i > \tau_j, \; i > j;$
\item $\tau_i \bmod N \neq \tau_j, \; i \neq j.$
\end{enumerate}
and $T$ be the set of all permutations of $\tau$. For the convenience we represent each tuple $\tau$ as $\tau = \tau_1 \cup \tau_2$ where $\tau_1 = \left \{ \tau_{1,i}\right\}_{i = 1}^{c_1}, \; \tau_2 = \left \{ \tau_{2,j}\right\}_{j = 1}^{c_2}$ and $\tau_{1,i} \leq N, \; 1 \leq i \leq c_1; \: \tau_{2,j} > N, \; 1 \leq j \leq c_2$. It's easy to observe that the set $\left \{ \tau_1, \; \tau_2 - N \right \}$ rearranged in increasing order of values coincides with the set $\left \{ 1, 2, \ldots, N \right \} $. Using elementary properties of the determinants, $\det S$ can be represented in the following form
\begin{equation} \label{eq:sys2236}
\det S = \sum_{\tau \in T}{ \det S_{\tau} \prod_{i = 1}^{c_1} {\mu_{N + \tau_{1,i}}} \prod_{j = 1}^{c_2} {\nu_{\tau_{2,j}}}},
\end{equation}
where $S_{\tau}$ - are the matrices whose $(\tau_{1,i})$-th columns  are constructed from $(\tau_{1,i})$-th columns of the matrix $S_1$ ($1 \leq i \leq c_1$) and $(\tau_{2,j} - N)$-th columns are constructed from $(\tau_{2,j} - N)$-th columns of the matrix $S_2$ ($1 \leq j \leq c_2$). 
It follows then that $S_{\tau}$ are \emph{paired Cauchy} matrices whose properties were investigated in \cite{heinig3}.
In order to calculate the determinants $\det S_{\tau}$,
consider two multi-sets of integers $\varkappa = \left \{ \varkappa_i\right\}_{i = 1}^{c_1}$ and $\bar{\varkappa} = \left \{ \bar{\varkappa}_j\right\}_{j = 1}^{c_2}$ defined in the following way:
\begin{enumerate}
\item $1 \leq \varkappa_{i} \leq N, \; 1 \leq i \leq c_1; \: 1 \leq \bar{\varkappa}_{j} \leq N, \; 1 \leq j \leq c_2;$
\item $\varkappa_i > \varkappa_j \; \text{if} \; i > j; \bar{\varkappa}_k > \bar{\varkappa}_l \; \text{if} \; k > l$;
\item $\varkappa \cap \bar{\varkappa} = \varnothing, \; \varkappa \cup \bar{\varkappa} = \left\{1, 2, \ldots, N \right\}$.
\end{enumerate}
Let $K$ be the set of all multi-sets $\varkappa$.
Using Laplace theorem and formulas~\eqref{eq:sys2233},~\eqref{eq:sys2234} the determinant of \emph{paired Cauchy} matrix corresponding to the tuple $\tau \in T$ is calculated as 
\begin{equation} \label{eq:sys2237}
\det S_{\tau} = \sum_{\varkappa \in K}{(-1)^{c_2 + \sum\limits_{k = 1}^{c_1}{\varkappa_k} + \sum\limits_{m = 1}^{c_1}{\tau_{1,m}}} \prod_{i = 1}^{c_1}{\mu_{\varkappa_i}} \prod_{j = 1}^{c_2}{\nu_{\bar{\varkappa}_j}}} \det S_{\varkappa} \det S_{\bar{\varkappa}},
\end{equation}
where
$$
\det S_{\varkappa} = \frac{\prod\limits_{1 \leq i < j \leq c_1} {(\xi_{\varkappa_i} - \xi_{\varkappa_j})} \prod\limits_{1 \leq i < j \leq c_1} {(\xi_{\varkappa_{N + i}} - \xi_{\varkappa_{N + j}})}}{\prod\limits_{\substack{1 \leq i \leq c_1 \\ 1 \leq j \leq c_1}} {(\xi_{\varkappa_i} - \xi_{\varkappa_{N + j}})}};
$$
$$
\det S_{\bar{\varkappa}} = \frac{\prod\limits_{1 \leq i < j \leq c_2} {(\xi_{\bar{\varkappa}_i} - \xi_{\bar{\varkappa}_j})} \prod\limits_{1 \leq i < j \leq c_2} {(\xi_{\bar{\varkappa}_{N + i}} - \xi_{\bar{\bar{\varkappa}}_{N + j}})}}{\prod\limits_{\substack{1 \leq i \leq c_2 \\ 1 \leq j \leq c_2}} {(\xi_{\bar{\varkappa}_i} - \xi_{\bar{\varkappa}_{N + j}})}}.
$$
Let's observe that summations over $T$ in~\eqref{eq:sys2236} and $K$ in~\eqref{eq:sys2237} produce unique combinations of the products
$$
P_{\tau} = \prod_{i = 1}^{c_1} {(\mu_{N + \tau_{1,i}} \; \mu_{\tau_{2,i} - N})} \prod_{j = 1}^{c_2} {(\nu_{\tau_{2,j}} \; \nu_{\tau_{1,j}})}
$$
expressed in terms of the tuple $\tau$. In order to unify and simplify indexation generated by $\tau$ and $\varkappa$ we introduce two $N$-tuples $\rho = \left \{ \rho_i \right \}_{i = 1}^N$ and $\bar{\rho} = \left \{ \bar{\rho}_i \right \}_{i = 1}^N$ defined as
\begin{enumerate}
\item $1 \leq \rho_{i} \leq 2N, \; 1 \leq i \leq N; \: 1 \leq \bar{\rho}_{j} \leq 2N, \; 1 \leq j \leq N;$
\item $\rho_i > \rho_j \; \text{if} \; i > j; \bar{\rho}_k > \bar{\rho}_l \; \text{if} \; k > l$;
\item $\rho \cap \bar{\rho} = \varnothing, \; \rho \cup \bar{\rho} = \left\{1, 2, \ldots, 2N \right\}$.
\end{enumerate}
Let $R$ represent the set of all permutations of $\rho$. Then it's easy to see that for each tuple $\tau$ there exists the set $\rho$ such that expressions for $P_{\tau}$ in terms of $\rho$ and $\bar{\rho}$ take the form
$$
P_{\rho} = \prod_{i = 1}^{N} {\mu_{\rho_i}} \prod_{j = 1}^{N} {\nu_{\bar{\rho}_j}}.
$$
After multiplying numerator and denominator of each term in~\eqref{eq:sys2237} corresponding to the tuple $\tau$ by 
$$
\prod\limits_{\substack{1 \leq i \leq c_1 \\ 1 \leq j \leq c_2}} {(\xi_{\varkappa_i} - \xi_{\bar{\varkappa}_{N + j}})} \prod\limits_{\substack{1 \leq i \leq c_2 \\ 1 \leq j \leq c_1}} {(\xi_{\bar{\varkappa}_i} - \xi_{\varkappa_{N + j}})}
$$
and substituting~\eqref{eq:sys2237} into~\eqref{eq:sys2236}, the expression for the coefficient $C_{\tau}$ by the term $P_{\tau}$ yields
$$
\begin{aligned}
C_{\tau} & = \prod\limits_{1 \leq i < j \leq c_1} {(\xi_{\varkappa_i} - \xi_{\varkappa_j})(\xi_{\varkappa_{N + i}} - \xi_{\varkappa_{N + j}})}
         \prod\limits_{1 \leq i < j \leq c_2} {(\xi_{\bar{\varkappa}_i} - \xi_{\bar{\varkappa}_j})(\xi_{\bar{\varkappa}_{N + i}} - \xi_{\bar{\varkappa}_{N + j}})} \\
				 & \times \prod\limits_{\substack{1 \leq i \leq c_1 \\ 1 \leq j \leq c_2}} {(\xi_{\varkappa_i} - \xi_{\bar{\varkappa}_{N + j}})} \prod\limits_{\substack{1 \leq i \leq c_2 \\ 1 \leq j \leq c_1}} {(\xi_{\bar{\varkappa}_i} - \xi_{\varkappa_{N + j}})} 
				 \left ( \prod\limits_{\substack{1 \leq i \leq N \\ N + 1 \leq j \leq 2N}} {(\xi_{i} - \xi_{j})}\right )^{-1}.
\end{aligned} 
$$
In terms of $\rho$ the expression for $C_{\tau}$ translates into
$$
C_{\rho} = \frac{\prod\limits_{1 \leq i < j \leq N} {(\xi_{\rho_i} - \xi_{\rho_j})}\prod\limits_{1 \leq i < j \leq N}{(\xi_{\bar{\rho}_{i}} - \xi_{\bar{\rho}_{j}})}}
				 {\prod\limits_{\substack{1 \leq i \leq N \\ N + 1 \leq j \leq 2N}} {(\xi_{i} - \xi_{j})}}.
$$
On the other hand, using Laplace theorem for the determinant $\Delta$ the following representation can be obtained 
\begin{equation} \label{eq:sys2238}
\Delta = \sum_{\rho \in R}{(-1)^{\sum\limits_{k = 1}^{N}{k} + \sum\limits_{m = 1}^{N}{\rho_{m}}} \prod_{i = 1}^{N}{\mu_{\rho_i}} \prod_{j = 1}^{N}{\nu_{\bar{\rho}_j}}{\prod\limits_{1 \leq i < j \leq N} {(\xi_{\rho_i} - \xi_{\rho_j})}\prod\limits_{1 \leq i < j \leq N}{(\xi_{\bar{\rho}_{i}} - \xi_{\bar{\rho}_{j}})}}}.
\end{equation}
Comparing~\eqref{eq:sys2238} with the previously obtained relations we conclude that $\det S$ and $\Delta$ are related by the formula
\begin{equation} \label{eq:sys2239}
\det S = (-1)^{\frac{N(N - 1)}{2}} \Delta \prod\limits_{\substack{1 \leq i \leq N \\ N + 1 \leq j \leq 2N}} {(\xi_{i} - \xi_{j})^{-1}}.
\end{equation}
As $\xi_{i} \neq \xi_{j}, \; 1 \leq i \leq N, \; N + 1 \leq j \leq 2N$ the assertion of the lemma follows.$\square$
\begin{thm}
  Let the sets of numbers $ \mu, \nu,
 \xi$ be such that $ \det S \neq 0$ then the transfer matrix-function $ W_A \left( \lambda
 \right)$ has the form~\eqref{eq:sys213} where

 \begin{center}
   $D_{11} \left( \lambda \right) = \left( -1 \right) ^ {N} \det\begin{bmatrix}
                                                            V_{N - 1} \left( \mu , \xi \right) & 0 \\
                                                            V_{N} \left( \nu , \xi \right) & \Lambda_N \\
                                                               \end{bmatrix}
  \Delta^{-1};$
  \end{center}
 \begin{center}
   $D_{12} \left( \lambda \right) = \left( -1 \right) ^ {N} \det\begin{bmatrix}
                                                            V_{N - 1} \left( \mu , \xi \right) & \Lambda_{N - 1} \\
                                                            V_{N}\left( \nu , \xi \right) & 0 \\
                                                          \end{bmatrix}
    \Delta^{-1};$
\end{center}

\begin{equation} \label{eq:sys224}
\end{equation}
 \begin{center}
   $D_{21} \left( \lambda \right) =  \det\begin{bmatrix}
                                                            V_{N} \left( \mu , \xi \right) & 0 \\
                                                            V_{N - 1} \left( \nu , \xi \right) & \Lambda_{N - 1} \\
                                                               \end{bmatrix}
  \Delta^{-1};$
  \end{center}
 \begin{center}
   $D_{22} \left( \lambda \right) =  \det\begin{bmatrix}
                                                            V_{N} \left( \mu , \xi \right) & \Lambda_{N} \\
                                                            V_{N - 1}\left( \nu , \xi \right) & 0 \\
                                                          \end{bmatrix}
    \Delta^{-1};$
\end{center}
and the matrix-function
$$
D(\lambda) = \left\{D_{ij}(\lambda)\right\}_{i,j = 1}^2
$$
is the \emph{basis solution} of the \textbf{IP Problem} with the degree $N - p - r$ where $p$ - is the number of indices $i_k, \; 1 \leq k \leq p$ for which $\mu_{i_k} = 0$ and $r$ - is the number of indices $j_k, \; 1 \leq k \leq r$ for which $\nu_{j_k} = 0$. Corresponding \emph{normalized basis solution} of the \textbf{IP Problem} is unique.
\end{thm}
\emph{Proof.} By virtue of the theorem conditions and \textbf{Lemma 2.2}, $\Delta \neq 0$.
Hence polynomials~\eqref{eq:sys224} make sense. From ~\eqref{eq:sys224} we find that 
\begin{equation} \label{eq:sys225}
\nu _i D_{k 1}( \xi _i ) + \mu _i D_{k 2}( \xi _i ) = 0, \; 1 \leq i \leq 2N, \; k = 1, 2. 
\end{equation}
Indeed, consider for example a combination 
$$
D_1(\lambda) = \nu_i D_{1 1}( \lambda ) + \mu_i D_{1 2}( \lambda ) 
$$
for some index $i: \; 1 \leq i \leq 2N$. It can be represented in the form
$$
D_1(\lambda) =  \left( -1 \right) ^ {N} \det\begin{bmatrix}
V_{N - 1} \left( \mu , \xi \right) & \mu_i \Lambda_{N - 1} \\
V_{N} \left( \nu , \xi \right) & \nu_i \Lambda_N \\
\end{bmatrix}
\Delta^{-1}.
$$
Setting $\lambda = \xi_i$ in the last expression, we see that $D_1(\xi_i)$ represents a determinant of the matrix whose $i$-th and last columns coincide. Thus $D_1(\xi_i) = 0, \; (1 \leq i \leq 2N)$. A combination corresponding to
$$
D_2(\lambda) =  \left( -1 \right) ^ {N} \det\begin{bmatrix}
V_{N} \left( \mu , \xi \right) & \mu_i \Lambda_{N} \\
V_{N - 1} \left( \nu , \xi \right) & \nu_i \Lambda_{N - 1} \\
\end{bmatrix}
\Delta^{-1}
$$
is treated analogously. So the equalities~\eqref{eq:sys221} are satisfied and $\left \{ D_{i j}(\lambda) \right \}_{i, j = 1}^2$ is the basis
solution of the \textbf{IP Problem}. We'll show now that the corresponding \emph{normalized basis solution} is unique. First, consider the case $\mu _i \neq 0 $ and $\nu _i \neq 0, \; 1 \leq i \leq 2N$. Assume that there exists another solution $\left \{ \tilde{D}_{i j}(\lambda) \right \}_{i, j = 1}^2$ of degree $N$ such that the equalities~\eqref{eq:sys221} are satisfied and the coefficients of the highest degree of the polynomial pairs $\left\{D_{1 1}(\lambda), \tilde{D}_{1 1}(\lambda)\right\}$ and $\left\{D_{2 2}(\lambda), \tilde{D}_{2 2}(\lambda)\right\}$ respectively, are equal to 1. Expressions~\eqref{eq:sys221} can be considered as two systems of $2N$ equations each with respect to the coefficients of the polynomials $D_{1 j}(\lambda)$ and $D_{2 j}(\lambda), \; j = 1, 2$.
We represent the polynomials $D_{1 1}(\lambda), D_{1 2}(\lambda)$ and $\tilde {D}_{1 1}(\lambda), \tilde {D}_{1 2}(\lambda)$ in the form
$D_{1 1}(\lambda) = \lambda ^{N} + \sum_{i = 1}^{N}{\lambda ^{N - i} d_{11}^{(i)}}$, \;
$D_{1 2}(\lambda) = \sum_{i = 1}^{N}{\lambda ^{N - i} d_{12}^{(i)}}$, \;
$\tilde {D}_{1 1}(\lambda) = \lambda ^{N} + \sum_{i = 1}^{N}{\lambda ^{N - i} \tilde {d}_{11}^{(i)}}$, \;
$\tilde {D}_{1 2}(\lambda) = \sum_{i = 1}^{N}{\lambda ^{N - i} \tilde {d}_{12}^{(i)}}$
and consider the systems
\begin{equation} \label{eq:sys226}
\nu _i D_{1 1}( \xi _i ) + \mu _i D_{1 2}( \xi _i ) = 0, \; 1 \leq i \leq 2N 
\end{equation}
and 
\begin{equation} \label{eq:sys227}
\nu _i \tilde {D}_{1 1}( \xi _i ) + \mu _i \tilde {D}_{1 2}( \xi _i ) = 0, \; 1 \leq i \leq 2N. 
\end{equation}
Subtracting corresponding equations in~\eqref{eq:sys226} from~\eqref{eq:sys227} we arrive at the system 
\begin{equation} \label{eq:sys228}
\nu _i \hat {D}_{1 1}( \xi _i ) + \mu _i \hat {D}_{1 2}( \xi _i ) = 0, \; 1 \leq i \leq 2N, 
\end{equation}
where
$$\hat {D}_{1 1}(\lambda) = \sum_{i = 1}^{N}{\lambda ^{N - i} (\tilde {d}_{11}^{(i)} - {d}_{11}^{(i)})},$$
$$\hat {D}_{1 2}(\lambda) = \sum_{i = 1}^{N}{\lambda ^{N - i} (\tilde {d}_{12}^{(i)} - {d}_{12}^{(i)})}.$$
According to the \textbf{Theorem 2.1} for the arbitrary polynomials $\hat {D}_{1 i}(\lambda), \; (i = 1, 2)$ of the degree less or equal $N - 1$ there exists a matrix $\hat{S}$ with the elements constructed from the sets $\left\{\nu_i\right\}, \;\left\{\mu_i\right\}$ and $\left\{\xi_i\right\}; \; (1 \leq i \leq 2N)$ and having the form~\eqref{eq:S} (hence $\hat{S} = S$) such that $\det S = 0$.
Again using \textbf{Lemma 2.2} we conclude that $\Delta = 0$ but this 
contradicts the condition of the theorem. Hence, $\tilde {d}_{11}^{(i)} - {d}_{11}^{(i)} = 0$ and $\tilde {d}_{12}^{(i)} - {d}_{12}^{(i)} = 0, \; i = 1, 2, \dots, N.$ The systems
\begin{equation} \label{eq:sys229}
\nu _i D_{2 1}( \xi _i ) + \mu _i D_{2 2}( \xi _i ) = 0, \; 1 \leq i \leq 2N 
\end{equation}
and 
\begin{equation} \label{eq:sys230}
\nu _i \tilde {D}_{2 1}( \xi _i ) + \mu _i \tilde {D}_{2 2}( \xi _i ) = 0, \; 1 \leq i \leq 2N 
\end{equation}
are considered analogously.\\
Let now $\mu_{i_k} = 0, \; k = 1, 2, \dots, p$ for some multi-index $i_1 , i_2, \dots i_p, \; 0 \leq p \leq 2N$ and $\nu_{j_m} = 0, m = 1, 2, \dots, r$ for some multi-index $j_1 , j_2, \dots j_r, \; 0 \leq r \leq 2N$. Consider first the system~\eqref{eq:sys226}. In this case polynomials $D_{1k}(\lambda), \; k = 1, 2$ can be represented as
\begin{equation} \label{eq:sys231}
D_{11}(\lambda) = B_{11} (\lambda) \prod_{k = 1}^p {(\lambda - \xi_{i_k})}, \;
D_{12}(\lambda) = B_{12} (\lambda) \prod_{m = 1}^r {(\lambda - \xi_{j_m})},
\end{equation}
where
$B_{1 1}(\lambda) = \lambda ^{N - p} + \sum_{i = 1}^{N - p}{\lambda ^{N - p - i} b_{11}^{(i)}}$, \;
$B_{1 2}(\lambda) = \sum_{i = 1}^{N - r}{\lambda ^{N - r - i} b_{12}^{(i)}}$.
Assume that there exists another solution $\tilde {D}_{1k}(\lambda), \; k = 1,2$ of the system~\eqref{eq:sys226} which can be represented as
\begin{equation} \label{eq:sys232}
\tilde {D}_{11}(\lambda) = \tilde{B}_{11} (\lambda) \prod_{k = 1}^p {(\lambda - \xi_{i_k})}, \;
\tilde {D}_{12}(\lambda) = \tilde{B}_{12} (\lambda) \prod_{m = 1}^r {(\lambda - \xi_{j_m})},
\end{equation}
where
$\tilde{B}_{1 1}(\lambda) = \lambda ^{N - p} + \sum_{i = 1}^{N - p}{\lambda ^{N - p - i} \tilde{b}_{11}^{(i)}}$, \;
$\tilde{B}_{1 2}(\lambda) = \sum_{i = 1}^{N - r}{\lambda ^{N - r - i} \tilde{b}_{12}^{(i)}}$. In case of \emph{normalized basis solution} the coefficients of the highest degree of the polynomials $B_{11} (\lambda)$ and $\tilde{B}_{11} (\lambda)$ equal one. Substituting~\eqref{eq:sys232} into~\eqref{eq:sys226} we arrive at two systems of equations with respect to the coefficients $\left \{ b_{11}^{(i)} \right \}_{i = 1}^{N - p}, \left \{ b_{12}^{(i)} \right \}_{i = 1}^{N - r}$ and $\left \{ \tilde {b}_{11}^{(i)} \right \}_{i = 1}^{N - p}, \left \{ \tilde {b}_{12}^{(i)} \right \}_{i = 1}^{N - r}$ respectively.
Subtracting corresponding equations of these systems we get 
\begin{equation} \label{eq:sys233}
\nu _j \sum_{i = 1}^{N - p}{\xi _j ^{N - p - i} (\tilde{b}_{11}^{(i)} - b_{11}^{(i)})} + \mu _j \sum_{i = 1}^{N - r}{\xi _j ^{N - r - i} (\tilde{b}_{12}^{(i)} - b_{12}^{(i)})} = 0, \; 1 \leq j \leq 2N.
\end{equation}
System~\eqref{eq:sys233} can have only trivial solutions otherwise it is required for the matrix of the coefficients to be singular which is equivalent to the condition $\det S = 0$ implying (according to \textbf{Lemma 2.2}) that $\Delta = 0$ but this contradicts the theorem's assumptions. Hence, $\tilde{b}_{11}^{(i)} = b_{11}^{(i)}, \; 1 \leq i \leq N - p$ and $\tilde{b}_{12}^{(i)} = b_{12}^{(i)}, \; 1 \leq i \leq N - r$.
\newline
Case of the polynomials $D_{2 j}(\lambda), \; j = 1, 2$ is considered analogously. $\square$
\newline
Now we summarize the properties of the polynomials $ D_{ij} \left( \lambda \right) \; \left( i, j = 1, 2  \right)$ under the condition $\Delta \neq 0$.
\begin{pro}
$$
\deg D_{11}(\lambda) = \deg D_{22}(\lambda) = N.
$$
\end{pro}
Assertion follows directly from~\eqref{eq:sys224}. It's easy to see that the coefficients of the highest degree of the polynomials $D_{11}(\lambda)$ and $D_{22}(\lambda)$ equal $\Delta$.
\begin{pro}
Coefficients of the polynomials $D_{ij}(\lambda), \; (i, j = 1, 2)$ do not depend on the absolute values of the parameters $\mu_i, \; \nu_i, \; (1 \leq i \leq 2N)$ but are determined up to the values of the ratios $\epsilon_i = \mu_i / \nu_i$ if $\mu_i \neq 0$ and $\nu_i \neq 0$.
\end{pro}
This follows from the equalities~\eqref{eq:sys221} where if $\mu_i \neq 0$ and $\nu_i \neq 0$, one can divide both sides by $\mu_i$ or $\nu_i$ without violating the equalities. 
\begin{pro}
The following equality is true
\begin{equation} \label{eq:sys234}
    D_{11} \left( \lambda \right) D_{22} \left( \lambda \right) - D_{12} \left( \lambda \right) D_{21} \left( \lambda \right) =
    \prod^{2N}_{k=1} \left( \lambda - \xi_k \right).
\end{equation}
\end{pro}
Indeed, formula~\eqref{eq:sys234} is the consequence of the expressions
$$
D_{11}(\xi _k) D_{22}(\xi _k) - D_{12}(\xi _k) D_{21}(\xi _k) = 0, \; k = 1, 2, \dots, 2N
$$
following from~\eqref{eq:sys221}.

\begin{pro}
 If $ i_1 , i_2 , \ldots i_p \left( 0 \leq
i_j \leq 2N, \; j = 1, 2, \ldots , p \right) $ are such that $
\mu_{i_j} = 0 \; \; \left( j = 1, 2, \ldots, p \right)$ then $ D_{11}
\left( \xi_{i_j} \right) = D_{21} \left( \xi_{i_j} \right) = 0$ and
vice versa, if $D_{11} \left( \xi_{i_j} \right) = 0 \left( D_{21}
\left( \xi_{i_j} \right) = 0 \right)$ then $D_{21} \left( \xi_{i_j}
\right) = 0 \left( D_{11} \left( \xi_{i_j} \right) = 0 \right)$ and
$ \mu_{i_j} = 0 \; $.
\end{pro}
Indeed, if one takes into account that simultaneous equalities $\mu_{i_j} = 0, \; \nu_{i_j} = 0$ are impossible then direct assertion follows from~\eqref{eq:sys221}. Now let's prove the inverse one. Assume that $D_{11}(\xi_{i_j}) = 0$. If $\nu_{i_j} \neq 0$, then $D_{k2}(\xi_{i_j}) \neq 0, \; (k = 1, 2)$ and from~\eqref{eq:sys221} it follows that $\mu_{i_j} = 0$ and $D_{21}(\xi_{i_j}) = 0$. Assuming that $\nu_{i_j} = 0$ implies the equalities $D_{k2}(\xi_{i_j}) = 0, \; (k = 1, 2)$ and the fact that the multiplicity of the root $\xi_{i_j}$ of the polynomial $D_{12}(\lambda)$ is greater than one. This contradicts the equality~\eqref{eq:sys234}.\\
The following property is proved similarly.   
\begin{pro}
If $ i_1 , i_2 ,
\cdots i_p \left( 0 \leq i_j \leq 2N, \; j = 1, 2, \cdots , p \right) $
are such that $ \nu_{i_j} = 0 \; \left( j = 1, 2, \cdots, p \right)$
then $ D_{22} \left( \xi_{i_j} \right) = D_{12} \left( \xi_{i_j}
\right) = 0$ and vice versa, if $D_{22} \left( \xi_{i_j} \right) = 0
\left( D_{12} \left( \xi_{i_j} \right) = 0 \right)$ then $D_{12}
\left( \xi_{i_j} \right) = 0 \left( D_{22} \left( \xi_{i_j} \right)
= 0 \right)$ and
$ \nu_{i_j} = 0 \; \left( j = 1, 2, \cdots, p \right)$
\end{pro}
\begin{pro}
The pairs of polynomials $\{D_{k1} \left(
\lambda \right),  D_{k2} \left( \lambda \right)\} \; \left( k = 1, 2
\right)$ do not have common roots.
\end{pro}
This follows from the fact that according to~\eqref{eq:sys234} polynomials $\{D_{1j} \left(
\lambda \right), \; (j = 1, 2)$ cannot have common roots other than $\xi_k$. But if $\xi_k$ - is the common root then \textbf{Property 2.4} implies that $\mu_{k} = 0$ and $\nu_{k} = 0$ which is impossible. The pair $\{D_{2j} \left(\lambda \right), \; (j = 1, 2)$ is considered analogously.\\\\  
Formulas~\eqref{eq:sys224} give a method of construction of the transfer
matrix-function of the dynamic system~\eqref{eq:TMFB} and basis solution of the corresponding interpolation problem~\eqref{eq:sys221} (\textbf {IP  Problem}). The \textbf{Properties 1 - 4}
are necessary for the existence of the functions $W_A \left( \lambda
\right)$ and $W_B \left( \lambda \right)$. \\
For the applications considered in this paper the \emph {inverse interpolation problem (\textbf{IIP Problem})} also plays an important role. The problem is formulated as follows:\\

\textbf{IIP Problem.} \emph{Given basis solution of \textbf{IP Problem} $\left \{X_{ij}(\lambda) \right \} _{i, j = 1}^2,$ find the sets $\{ \mu _i\}, \{ \nu _i\}, \{\xi _i \}, \; 1 \leq i \leq 2N$ satisfying the equalities~\eqref{eq:sys222}}.

As  it was noted in the \textbf{Property 2.2} the mapping $
\{D_{ij} \left( \lambda \right)\} _{i, j = 1}^2 \rightarrow \{ \mu, \nu, \xi \}$
is not unique. Given the polynomials $ D_{ij} \left( \lambda \right)
\; \left( i, j = 1, 2
 \right)$ the sets $\{ \mu, \nu, \xi \}$ can be
restored only up to the ratios $ \mu_i / \nu_i \left(
i = 1, 2, \cdots, 2N \right)$. So the mapping $ \{ \mu, \nu, \xi \}
\rightarrow \{D_{ij} \left( \lambda \right)\}$ under some
conditions is surjective. The following theorem formulates these conditions and gives the solution of \textbf {IIP Problem}. 

\begin{thm}
 Let the polynomials $ D_{ij} \left(
\lambda \right) \; \left( i, j = 1, 2
 \right)$ be such that

\indent \emph{a)  $\deg\{D_{11} \left( \lambda \right)\} =
\deg\{D_{22} \left( \lambda \right)\} = N,$}

\indent   \hspace{12 pt}  $\deg\{D_{12} \left( \lambda \right)\} \leq
N - 1, \;
\deg\{D_{21} \left( \lambda \right)\} \leq N - 1;$\\
\indent \emph{b) pairs of the polynomials $ \{ D_{k1}\left( \lambda
\right), D_{k2}\left( \lambda \right)\}, \left( k = 1, 2 \right)$ do
not have common roots;}\\
\indent \emph{c) polynomial $ D\left( \lambda \right) = D_{11}
\left( \lambda \right) D_{22} \left( \lambda \right) - D_{12} \left(
\lambda \right) D_{21} \left( \lambda \right)$ has simple roots.}\\
Then the sets $ \{ \mu, \nu, \xi \}$ can be restored up to the ratios
 $ \mu_i / \nu_i $ $\left( i = 1, 2, \cdots,
2N
\right)$ and the corresponding matrix $S$ is non-singular.
\end{thm}
\emph{Proof}. Indeed, given the polynomials $D_{ij} \left( \lambda \right)\;\left(
i, j = 1, 2
 \right)$ satisfying the requirements a) - c) of the Theorem, let $\hat{\xi_k} \; \left( k = 1, 2, \cdots, 2N\right)$
be the simple roots of the polynomial $ D\left( \lambda \right)$. Consider
the following cases:\\

\indent 1) for all $k = 1, 2, \ldots, 2N$ we have $D_{ij}\left( \hat{\xi_k} \right) \neq 0, \; \left( i, j = 1, 2 \right)$;\\

\indent 2) for some $i_1,i_2, \cdots, i_p : 1 \leq i_j \leq 2N, \;
\left(j = 1, 2, \cdots, p \right)$ one has $ D_{k1}\left(
\hat{\xi_{i_j}} \right) =
0, \; \left( k = 1, 2 \right)$;\\
\indent 3) for some $l_1,l_2, \cdots, l_m : 1 \leq l_s \leq 2N, \;
\left(s= 1, 2, \cdots, m \right)$ one has $ D_{k2}\left(
\hat{\xi_{l_s}} \right)
= 0, \; \left( k = 1, 2 \right)$.\\

In the case 1) let $ \hat{\mu_i} \left(1 \leq i \leq 2N \right)$ be a
set of arbitrary non-zero numbers. Define
\begin{equation} \label{eq:sys235}
    \hat{\nu_i} = \hat{\mu_i}\frac{D_{12} \left( \hat{\xi_i} \right)}{D_{11} \left( \hat{\xi_i}
    \right)} = \hat{\mu_i}\frac{D_{22} \left( \hat{\xi_i} \right)}{D_{21} \left( \hat{\xi_i}
    \right)}, \; 1 \leq i \leq 2N.
\end{equation}

In the case 2) we put $ \hat{\mu}_{i_j} = 0 \; \left(j = 1, 2, \cdots, p
\right)$ and let $\hat{\mu}_{k} \; \left( k \neq i_j \right) $ be
arbitrary non-zero numbers. Choose $ \hat{\nu}_{i_j} \; \left(j = 1,
2, \cdots, p \right)$ as arbitrary non-zero numbers and let
\begin{equation} \label{eq:sys236}
    \hat{\nu}_k = \hat{\mu}_k\frac{D_{12} \left( \hat{\xi_i} \right)}{D_{11} \left( \hat{\xi_i}
    \right)} = \hat{\mu}_k\frac{D_{22} \left( \hat{\xi_i} \right)}{D_{21} \left( \hat{\xi_i}
    \right)}, \; 1 \leq i \leq 2N, \; k \neq i_j.
\end{equation}

In the case 3) we put $ \hat{\nu}_{l_s} = 0 \; \left(l = 1, 2, \cdots, m
\right)$ and let $\hat{\nu}_{k} \; \left( k \neq l_s \right) $ be
arbitrary non-zero numbers. Choose $ \hat{\mu}_{l_s} \; \left(l = 1,
2, \cdots, m \right)$ as arbitrary non-zero numbers and let
\begin{equation} \label{eq:sys237}
    \hat{\mu_k} = \hat{\nu}_k\frac{D_{11} \left( \hat{\xi_i} \right)}{D_{12} \left( \hat{\xi_i}
    \right)} = \hat{\nu}_k\frac{D_{21} \left( \hat{\xi_i} \right)}{D_{22} \left( \hat{\xi_i}
    \right)}, \; 1 \leq i \leq 2N, \; k \neq l_s.
\end{equation}

Then the sets $ \{\hat{\xi}\}, \; 
\{\hat{\mu}\}, \; \{\hat{\nu}\}$ solve our problem and the corresponding matrix $S$ is non-singular. $\square$

\section{Singular solutions of non-linear integrable differential equations}
Material of this section is based on the results obtained in a number of papers (see for example \cite{lev5} - \cite{lev7}) where the method of operator identities was successfully applied to obtaining explicit solutions of some NIDE using the inverse spectral problem approach. Later this approach was extended to obtain more general classes of solutions (see \cite{sasa1, sasa5, sasa6}). The majority of the results of this section are not new and can be interpreted as scalar analogues of the formulas from \cite{sasa2, sasa3, sasa4}. At the same time, the simplicity of our
 special case allows to perform more thorough investigation of the properties of the solutions and obtain more detailed information of their behavior. We consider the following NIDEs
\begin{equation} \label{eq:SHG}
\frac {\partial ^2 \phi (x, t)}{\partial x \partial t} = 4 \sinh \phi (x, t) \quad \text{sinh-Gordon equation (SHG) };
\end{equation}
\begin{equation} \label{eq:MKDV}
\begin{split}
\frac {\partial \psi (x, t)}{\partial t} = - \frac{1}{4} \frac {\partial ^3 \psi (x, t)}{\partial x ^3} +
\frac{3}{2} | \psi (x, t) |^2 \frac {\partial \psi (x, t)}{\partial x}
\\
\text{modified Korteweg - de Vries equation (MKdV)};
\end{split}
\end{equation}

\begin{equation} \label{eq:NSE}
\begin{split}
\frac {\partial \rho (x, t)}{\partial t} = \frac{\imath}{2} \left [\frac {\partial ^2 \rho (x, t)}{\partial x ^2} -
2 | \rho (x, t) |^2 \rho (x, t) \right ]
\\
\text{non-linear Schr\" {o}dinger equation (NSE)}.
\end{split}
\end{equation}
The method of inverse spectral problem as opposed to the method of inverse scattering, allows to weaken the requirement of the solution regularity and investigate solutions with singularities (inverse scattering approach requires the regularity of the solutions on the axis $x \in (-\infty, \infty)$ while inverse spectral problem method requires regularity of the solutions on semi-axis $x \in (0, \infty)$). Below we sketch the results obtained in this way following \cite{lev3} and \cite{dis}.\\

\subsection{Explicit solutions of NIDE}
To the equations~\eqref{eq:SHG}-~\eqref{eq:NSE} we associate the following linear system of differential equations:
\begin{equation} \label{eq:sys34}
\frac {\partial W}{\partial x} = \imath z H(x, t) \; W, \: W(0, t, z) = I_2,
\end{equation} 
where 
\begin{equation} \label{eq:sys35}
H(x, t) = \begin{bmatrix}
  0 & exp[\xi (x, t) - \xi (0, t)] \\
  exp[\xi (0, t) - \xi (x, t)] & 0,
 \end{bmatrix}.
\end{equation} 
$\xi (x, t)$ - is the solution of either of the equations~\eqref{eq:SHG}-~\eqref{eq:NSE} and $I_2$ - is $2 \times 2$ identity matrix. In further considerations we'll always, unless specifically stated, assume that the function $\xi (x, t), \; 0 \leq x \leq \infty, \; 0 \leq t \leq \infty$ is real valued. In this case equalities ~\eqref{eq:sys34},~\eqref{eq:sys35} represent self-adjoint canonical system of differential equations for which Weyl-Titchmarsh function $v(t,z)$ is defined by the following inequality
\begin{equation} \label{eq:sys36}
\int_0^\infty {[1 \; \imath v^* (t, z) ] W^* (x, t, z) [J \; H(x, t) ] W(x, t, z)
\begin{bmatrix}
  1  \\
  -\imath \; v(t, z)
 \end{bmatrix}dx \; < \infty },
\end{equation}
where $\Im z > 0$ and 
$$
J = \begin{bmatrix}
  0 & 1 \\
  1 & 0
 \end{bmatrix}.
$$

In case of rational function $v(t,z)$, explicit solutions of non-linear equations can be constructed. We consider a special class $\mathcal{P}$ of functions $v(z)$ satisfying the following conditions
\begin{itemize}
\item {Function $v(z)$ is rational with poles in the lower half plain $\Im z < 0$, and $\Im v(z) > 0 \; \text {if} \; \Im z \geq 0; \; \Re v(z) = 0 \; \text {when} \; \Re z = 0$};
\item {$\lim _{z \to \infty} {v(z)} = 0; \; -\imath v(0) > 0$}.
\end{itemize}
\begin{Rk}
Function $v(z)$ belongs to Nevanlinna class , i. e. satisfies the conditions
$$
\Im v(z) > 0, \; \text{if} \; \Im z > 0.
$$
\end{Rk}
Below, closely following \cite{lev2}, we describe the procedure of construction of the explicit solutions which consists of the several steps.\\
\textbf{Step 1}. Let $v_0 (z) \equiv v(0, z) \in \mathcal{P}$, and 
\begin{equation} \label{eq:sys37}
v_0 (z) = \imath \prod _{k = 1}^N{\frac {z + \imath \gamma _{k, 0}} {z + \imath \alpha _{k, 0}}}; \quad \gamma _{k, 0} \neq \gamma _{j, 0}, \quad \alpha _{k, 0} \neq \alpha _{j, 0}; \quad k \neq j,
\end{equation}
i.e. function $v_0 (z)$ is rational Nevanlinna-type function with distinct sets of zeros $ \{- \imath \gamma_{k, 0}\}$ and poles $ \{- \imath \alpha_{k, 0}\}; \quad 1 \leq k \leq N$.
On the sufficiently small interval $0 \leq t \leq T$ function $v(z, t)$ has the form
\begin{equation} \label{eq:sys38}
v (z, t) = \imath \prod _{k = 1}^N{\frac {z + \imath \gamma _k (t)} {z + \imath \alpha _k (t)}}; \quad \gamma _k (0) = \gamma _{k, 0}, \quad \alpha _k (0) = \alpha _{k, 0}.
\end{equation}
\textbf{Step 2}. We construct the polynomial
\begin{equation} \label{eq:sys39}
Q (z) = \frac {(-1)^N}{2} [P_1 (z, t) P_2 (-z, t) + P_1 (-z, t) P_2 (z, t)], 
\end{equation} 
where 
\begin{equation} \label{eq:sys391}
P_1 (z, t) = \prod _{k = 1} ^N {[z - \alpha _k (t)]}; \quad P_2 (z, t) = \prod _{k = 1} ^N {[z - \gamma _k (t)]}.
\end{equation}
Assume that the roots $\omega _k \; (1 \leq k \leq 2N)$ of the polynomial $Q (z)$ are such that $\omega _k \neq \omega _i$ if $k \neq i$. It has been proven in \cite{lev8} that the numbers $\omega _k \: (1 \leq k \leq 2N)$ are integrals of motion (do not depend on $t$). The following theorem is true
\begin{thm}
The following evolution ($t$-dependance) formulas hold
\begin{equation} \label{eq:sys310}
\frac {P_i (\omega_k , t)}{P_i (-\omega_k , t)} = \frac {P_i (\omega_k , 0)}{P_i (-\omega_k , 0)} \exp {(-2t \Theta (\omega_k))}; \quad i = 1, 2; \quad 1 \leq k \leq 2N,
\end{equation} 
where $P_i(t)$ are expressed via $\alpha_k(t)$ and $\gamma_k(t)$ in~\eqref{eq:sys391}, $\omega_i \; (1 \leq i \leq 2N)$ are zeros of $Q(z)$ and
\[ \Theta (x) =
  \begin{cases}
    1/x       & \; \text{in case of SHG equation}; \\
    -x^3  & \; \text{in case of MKdV equation}; \\
		\imath x^2  & \; \text{in case of NSE equation}. \\
  \end{cases}
\]
\end{thm}
From ~\eqref{eq:sys39} it follows that 
\begin{equation} \label{eq:sys311}
\frac {P_1 (\omega_k , 0)}{P_1 (-\omega_k , 0)} = -\frac {P_2 (\omega_k , 0)}{P_2 (-\omega_k , 0)}, \quad 1 \leq k \leq 2N.
\end{equation} 

\begin{Dn}
 Given a set of quantities $Y = \left \{y_i \right \}_{i = 1}^N$ and ordered sets $\mathcal{I}_j = \left \{i_k \right \}_{k = 1}^j$ of indexes such that $i_m \neq i_n, \: m \neq n; \: i_m > i_n, \: m > n; \: 1 \leq k \leq j$, the elementary symmetric form $\sigma _j (Y)$ of order $j$ is defined as 
$$
\sigma _j (Y) = \sum _{i_k \in \mathcal{I}_j} \prod_{k = 1}^{j}{y_{i_k}}, \; 1 \leq j \leq N; \quad \sigma _0 (Y) = 1.
$$
\end{Dn}
Equalities~\eqref{eq:sys311} can be written as two systems of linear equations with respect to the symmetric forms $\sigma (A(t))$ and $\sigma (G (t))$ where $A(t) = \left \{\alpha _k (t) \right \}_{k = 1}^N$ and $G(t) = \left \{\gamma _k (t) \right \}_{k = 1}^N$:
\begin{equation} \label{eq:sys3111}
\sum_{j = 0}^N{(-1)^{N + j} \omega_k^j \sigma_j (A(t))} = \frac {P_1 (\omega_k , 0)}{P_1 (-\omega_k , 0)} \exp {(-2t \Theta (\omega_k))} \sum_{j = 0}^N{\omega_k^j \sigma_j (A(t))},
\end{equation}
and
\begin{equation} \label{eq:sys3112}
\sum_{j = 0}^N{(-1)^{N + j} \omega_k^j \sigma_j (G(t))} = \frac {P_2 (\omega_k , 0)}{P_2 (-\omega_k , 0)} \exp {(-2t \Theta (\omega_k))} \sum_{j = 0}^N{\omega_k^j \sigma_j (G(t))},
\end{equation}
where $1 \leq k \leq 2N$. It's easy to see that systems~\eqref{eq:sys3111},~\eqref{eq:sys3112} have unique solutions.\\ 
\textbf{Step 3}. Solving~\eqref{eq:sys3111} and~\eqref{eq:sys3112} with respect to $\sigma (A(t))$ and $\sigma (G (t))$ and substituting results in 
\begin{equation} \label{eq:sys312}
v(t, z) = \imath \frac {\sum _{k = 0}^N {\imath ^{N - k} z^k \sigma_{N - k} (G (t))}}{\sum _{k = 0}^N {\imath ^{N - k} z^k \sigma_{N - k} (A (t))}}
\end{equation}
we obtain an explicit representation for the evolution ($t$-dependence) of the Weyl-Titchmarsh function.
In \cite{lev2} it has been proven
\begin{thm}
If $v(z) \in \mathcal{P}$, then $v(t, z) \in \mathcal{P}$ and the number of zeros and poles, including their multiplicities, is preserved.
\end{thm}  

\textbf{Step 4}. Introduce in $L_2 (0, \zeta)$ an operator
\begin{equation} \label{eq:sys313}
S_\zeta \; g = 2 \; g(x) + \int_0 ^\zeta {g(u) \; K(x - u, t) \; du},
\end{equation} 

where
\begin{equation} \label{eq:sys314}
\begin{split}
r(t, z) = \frac {1}{\pi}[\Im v(t, z) - 1], \\
K(x, z) = \int _{-\infty}^\infty {e ^{\imath z x } \; r(t, z) \; dz}.
\end{split}
\end{equation}
The inverse $S_\zeta ^{-1}$ admits the representation 
\begin{equation} \label{eq:sys315}
S_\zeta ^{-1}\; g = \frac {1}{2} \; g(x) + \int_0 ^\zeta {g(u) \; \Gamma _\zeta (x, u, t) \; du}.
\end{equation} 
\textbf{Step 5}. The solution of the equations (3.1)-(3.3) is represented as
\[ \xi (x, t) =
  \begin{cases}
    8 \int _x ^\infty {\Gamma _{2s} (2s, 0, t)ds} & \; \text{in case of SHG equation }; \\
    -4 \Gamma _{2x} (2x, 0, t)  & \; \text{in case of MKdV and NSE equation }. \\
  \end{cases}
\]
In case of SHG equation the solution $\xi (x, t)$ can also be written as
\begin{equation} \label{eq:sys316}
\xi (x, t) = -\sinh^{-1} (2 \frac {\partial}{\partial t}\Gamma _{2x} (2x, 0, t)).
\end{equation}
Let's note that $\sinh^{-1}(x) = \ln(x + \sqrt{1 + x^2})$.
\begin{Rk}
Although the above procedure has been developed for the case when the system~\eqref{eq:sys34},~\eqref{eq:sys35} is self-adjoint canonical (function $\xi(x, t)$ is real-valued), the explicit nature of the solution allows its analytical prolongation into the complex domain.
\end{Rk}
Equalities~\eqref{eq:sys311},~\eqref{eq:sys3111},~\eqref{eq:sys3112} suggest that the dynamics (dependence on $t$) of the functions $\alpha_k(t)$ and $\gamma_k(t), \; (1 \leq k \leq N)$ is quite similar. This allows to express Weyl-Titchmarsh function and the solution of NIDE in terms of only $\left \{\omega_i \right \}_{k = i}^{2N}$ and $\left \{\alpha_{k,0} \right \}_{k = i}^N$.
Consider the case when the set $A(\alpha_0) = \left \{\alpha_{k,0} \right \}_{k = i}^N$
is symmetric with respect to the real axis, then the roots of the polynomial ~\eqref{eq:sys39} $\Omega  = \left \{\omega_k \right \}_{k = i}^{2N}$ are symmetric with respect to both real and imaginary axis. In further considerations by $\Omega (\omega)$ we denote the set $\left \{\omega_k \right \}_{k = i}^N$ where
 $\Re \omega_k > 0, \; 1 \leq k \leq N$ and $\omega_i \neq \omega_k, \; i \neq k$ and assume $\omega_{N + k} = -\omega_k, \; 1 \leq k \leq N$.
Application of~\eqref{eq:sys313}-~\eqref{eq:sys315} gives the following representation for the function $\Gamma _{2x} (2x, 0, t)$: 
\begin{equation} \label{eq:Gamma}
\begin{aligned}
    \Gamma _{2x} (2x, 0, t) = {} &\frac {1}{2} 
[\psi_1 (x, t) \; \cdots \; \psi_{N} (x, t) \; \psi_1^{-1} (x, t) \; \cdots \; \psi_{N}^{-1} (x, t)] \\ 
& S^{-1}(x, t) \;
[\underbrace{1 \qquad \cdots \qquad 1}_N \; \underbrace{0 \qquad \cdots \qquad 0}_N ]^T,
\end{aligned}
\end{equation}
where 

\begin{equation} \label{eq:sys321}
S(x, t) = 
\begin{bmatrix}
          \frac{1}{\omega_1 + \alpha_{1,0}} & \cdots & \frac{1}{\omega_N + \alpha_{1,0}} & \frac{1}{\alpha_{1,0} - \omega_1} & \cdots & \frac{1}{\alpha_{1,0} - \omega_N} \\
          \cdots & \cdots & \cdots & \cdots & \cdots & \cdots\\
          \frac{1}{\omega_1 + \alpha_{N,0}} & \cdots & \frac{1}{\omega_N + \alpha_{N,0}} & \frac{1}{\alpha_{N,0} - \omega_1} & \cdots & \frac{1}{\alpha_{N,0} - \omega_N} \\
 & & & & & \\
          \frac{\psi_1 (x, t)}{\omega_1 - \alpha_{1,0}} & \cdots & \frac{\psi_N (x, t)}{\omega_N - \alpha_{1,0}} & \frac{-1}{\psi_1 (x, t) \left( \alpha_{1,0} + \omega_1 \right)} & \cdots & \frac{-1}{\psi_N (x, t) \left( \alpha_{1,0} + \omega_N \right)} \\
          \cdots & \cdots & \cdots & \cdots & \cdots & \cdots\\
          \frac{\psi_1 (x, t)}{\omega_1 - \alpha_{N,0}} & \cdots & \frac{\psi_N (x, t)}{\omega_N - \alpha_{N,0}} & \frac{-1}{\psi_1 (x, t) \left( \alpha_{N,0} + \omega_1 \right)} & \cdots & \frac{-1}{\psi_N (x, t) \left( \alpha_{N,0} + \omega_N \right)}\\
\end{bmatrix}
\end{equation}
and $\psi_k (x, t) = \exp {[2(\omega_k x + \Theta (\omega_k) t)]}$. Performing calculations on the right hand side of ~\eqref{eq:Gamma} (for the details see \cite{dis}) one obtains the following expression for the function $\Gamma _{2x} (2x, 0, t)$
\begin{equation} \label{eq:Gamma1}
\Gamma _{2x} (2x, 0, t) = \frac {(-1)^N}{2} \frac {\Delta_1 (x, t)}{\Delta_2 (x, t)},
\end{equation}
where 
\begin{equation} \label{eq:Gamma2}
\Delta_1 (x, t) =
det \begin{bmatrix}
		1 & \cdots & 1 & 1 & \cdots & 1 \\
		\omega_1 & \cdots & \omega_N & -\omega_1 & \cdots & -\omega_N \\
		\cdots & \cdots & \cdots & \cdots & \cdots & \cdots \\
		\omega_1^{N - 2} & \cdots & \omega_N^{N - 2} & (-1)^{N - 2} \omega_1^{N - 2} & \cdots & (-1)^{N - 2}\omega_N^{N - 2} \\
		\psi_1(x, t) & \cdots & \psi_N(x, t) & \frac {1}{\psi_1 (x, t)} & \cdots & \frac {1}{\psi_N (x, t)}(x, t) \\
		\omega_1 \psi_1 (x, t) & \cdots & \omega_N \psi_N (x, t) & \frac {-\omega_1}{\psi_1 (x, t)} & \cdots & \frac {-\omega_N}{ \psi_N (x, t)}\\
		\cdots & \cdots & \cdots & \cdots & \cdots & \cdots \\
		\omega_1 ^N \psi_1 (x, t) & \cdots & \omega_N ^N \psi_N (x, t) & \frac{(-1)^N \omega_1 ^N}{ \psi_1 (x, t)} & \cdots & \frac{(-1)^N\omega_N ^N}{ \psi_N (x, t)}\\
		\end{bmatrix},
\end{equation}
\begin{equation} \label{eq:Gamma3}
\Delta_2 (x, t) =
det \begin{bmatrix}
		1 & \cdots & 1 & 1 & \cdots & 1 \\
		\omega_1 & \cdots & \omega_N & -\omega_1 & \cdots & -\omega_N \\
		\cdots & \cdots & \cdots & \cdots & \cdots & \cdots \\
		\omega_1^{N} & \cdots & \omega_N^{N} & (-1)^{N} \omega_1^{N} & \cdots & (-1)^{N}\omega_N^{N} \\
		\psi_1 (x, t) & \cdots & \psi_N  (x, t) & \frac {1}{\psi_1 (x, t)} & \cdots & \frac {1}{\psi_N (x, t)} \\
		\omega_1 \psi_1 (x, t) & \cdots & \omega_N \psi_N (x, t) & \frac {-\omega_1}{\psi_1 (x, t)} & \cdots & \frac {-\omega_N}{ \psi_N (x, t)}\\
		\cdots & \cdots & \cdots & \cdots & \cdots & \cdots \\
		\omega_1 ^{N - 2} \psi_1 (x, t) & \cdots & \omega_N ^{N - 2} \psi_N (x, t) & \frac {(-1)^{N - 2} \omega_1 ^{N - 2}}{ \psi_1 (x, t)} & \cdots & \frac {(-1)^{N - 2}\omega_N ^{N - 2}}{ \psi_N (x, t)}\\
		\end{bmatrix}.
\end{equation} 
The following theorems summarize the above.
\begin{thm}
 Let $\Omega = \left \{ \omega_k \right \}_{k = 1}^N; \; \A_0 = \left \{ \alpha_{k, 0} \right \}_{k = 1}^N$ be two sets of numbers such that $\omega_i \neq \omega_k, \; i \neq k; \; \Re \omega_k > 0$. Then the solution of MKdV and NSE equations can be represented as 
\begin{equation} \label{eq:sys325}
\xi (x, t) = 2 (-1)^{N -1} \frac{\Delta_1(x, t)}{\Delta_2(x, t)} 
\end{equation}
where $\Delta_k(x, t), \; k = 1, 2$ are defined by~\eqref{eq:Gamma2} and ~\eqref{eq:Gamma3} and 
\[ \psi_k (x, t) =
  \begin{cases}
    \exp[2 (\omega_k x -\omega_k^3 t - C_k)] \; \text{in case of MKdV equation;}\\
    \exp[2 (\omega_k x +\imath \omega_k^2 t - C_k)] \; \text{in case of NSE equation}. \\
  \end{cases}
\]
with $C_k = \frac {1}{2} \ln (\prod _{i = 1}^N |\frac {\omega_j - \alpha_{i, 0}} {\omega_j + \alpha_{i, 0}}|)$.
\end{thm}
Let's introduce the notations
\begin{equation}\label{eq:Delta1}
\delta_1 (x, t) = \det \Big\{\omega_j ^{N} \frac{\partial ^k}{\partial t^k} \cosh \chi_j(x, t)\Big\}_{0 \leq j, k \leq N - 1},
\end{equation}
\begin{equation}\label{eq:Delta2}
\delta_2 (x, t) = \det \Big\{\omega_j ^{N} \frac{\partial ^k}{\partial t^k} \sinh \chi_j(x, t)\Big\}_{0 \leq j, k \leq N - 1},
\end{equation}
where $\chi_j = \omega_j x + t / \omega_j -\frac {1}{2} \ln (\prod _{i = 1}^N |\frac {\omega_j - \alpha_{i, 0}} {\omega_j + \alpha_{i, 0}}|)$.
\begin{thm}
 Let $\Omega = \left \{ \omega_k \right \}_{k = 1}^N; \; \A_0 = \left \{ \alpha_{k, 0} \right \}_{k = 1}^N$ be two sets of numbers such that $\omega_i \neq \omega_k, \; i \neq k; \; \Re \omega_k > 0$ and each of the sets $\Omega, \; \A_0$ is symmetric with respect to the real axis. Then the solution $\xi (x, t)$ of SHG equation is represented as
\begin{equation} \label{eq:SH}
\xi (x, t) = 2 \ln \Big|\frac {\delta_1 (x, t)} {\delta_2 (x, t) } \Big|.
\end{equation}
\end{thm}
We illustrate formulas~\eqref{eq:sys325} and~\eqref{eq:SH} on some simple examples.
\begin{Ee}
Let $N = 1$ and $\omega_1 = \bar{\omega}_1 \equiv \omega, \; \alpha_{1, 0} = \bar{\alpha}_{1, 0} \equiv \alpha_0$, then according to~\eqref{eq:Delta1} and~\eqref{eq:Delta2}
\[ \delta_1 (x, t) =
  \begin{cases}
    \sinh (\omega x + \frac {t}{\omega} - \frac {1}{2} \ln | \frac {\omega - \alpha_0}{\omega + \alpha_0} |), & \; \text{if} \: \omega > \alpha_0;\\
    \cosh (\omega x + \frac {t}{\omega} - \frac {1}{2} \ln | \frac {\omega - \alpha_0}{\omega + \alpha_0} |), & \; \text{if} \: \omega < \alpha_0.\\
  \end{cases}
\]
 and 
\[ \delta_2 (x, t) =
  \begin{cases}
    \cosh (\omega x + \frac {t}{\omega} - \frac {1}{2} \ln | \frac {\omega - \alpha_0}{\omega + \alpha_0} |) & \; \text{if} \: \omega > \alpha_0\\
    \sinh (\omega x + \frac {t}{\omega} - \frac {1}{2} \ln | \frac {\omega - \alpha_0}{\omega + \alpha_0} |) & \; \text{if} \: \omega < \alpha_0\\
  \end{cases}
\]
It's easy to verify that the function $\xi (x, t) = 2 \ln |\frac {\delta_1 (x, t)} {\delta_2 (x, t) } |$ satisfies SHG equation~\eqref{eq:SHG}.
\end{Ee}
\begin{Ee}
Let $N = 1$, then from~\eqref{eq:Gamma} we deduce that if $\omega > \alpha_0$ then the function
$$
\xi (x, t) = 2 \Re \omega \exp \imath (\Im \chi (x, t)) / \sinh (\Re \chi (x, t))
$$
satisfies mKdV equation with
$$\chi (x, t) = 2 (\omega x - \omega ^3 t - \frac {1}{2} \ln | \frac {\omega - \alpha_0}{\omega + \alpha_0} |),
$$
and NSE with
$$\chi (x, t) = 2 (\omega x + \imath \omega ^2 t - \frac {1}{2} \ln | \frac {\omega - \alpha_0}{\omega + \alpha_0} |).
$$
If $\omega < \alpha_0$ then $\sinh (\Re \chi (x, t))$ is replaced by $\cosh (\Re \chi (x, t))$.
\end{Ee}
\subsection{Dynamic systems and associated inverse problems for NIDE}
In this paragraph we refer to the dynamic system corresponding to the $S$-node defined by~\eqref{eq:identity} and associated matrix $S$ of type~\eqref{eq:S}. Matrix-function $S(x, t)$~\eqref{eq:sys321} is a special case of the matrix $S$. Indeed, the equalities
\begin{equation} \label{eq:sys}
\begin{aligned}
& a_k = 1, \; a_{N + k} = 0; \; 1 \leq k \leq N; \\
& c_k = 0, \; c_{N + k} = 1; \;  \leq k \leq N; \\
& d_k = \psi_k (x, t), \; d_{N + k} = \psi_k^{-1} (x, t); \; 1 \leq k \leq N; \\
& g_k = \alpha_{k, 0}, \; g_{N + k} = -\alpha_{k, 0}; \; 1 \leq k \leq N; \\
& h_k = \omega_k, \; h_{N + k} = -\omega_k, \; 1 \leq k \leq N; \\
&  b_k = 1, \; 1 \leq k \leq 2N; \\
\end{aligned}
\end{equation}
map matrix $S$ onto $S(x, t)$. Results obtained for the matrix $S$ in the previous section and the fact that matrix-function $S(x, t)$ is a special case of the matrix $S$ allow us to make a connection between dynamic systems and NIDE. In this section we formulate and solve some of the related problems.\\
For the convenience we present here the definitions for the matrices $A, B, \Pi_1 , \Pi_2$ from matrix identity~\eqref{eq:identity} in this special case.\\
$$
A=diag\{\alpha_{1, 0} , \ldots , \alpha_{N, 0}, -\alpha_{1, 0} , \ldots , -\alpha_{N, 0} \} ,
$$\\
$$
B=diag\{-\omega_{1} , \ldots , -\omega_{N}, \omega_{1} , \ldots , \omega_{N} \},
$$
\\
$$
        \Pi ^T _1 = \begin{bmatrix}
                      1 & \ldots & 1 & 0 & \ldots & 0 \\
                      0 & \ldots & 0 & 1 & \ldots & 1 \\
                    \end{bmatrix},
$$\\
$$										
        \Pi ^T _2 = \begin{bmatrix}
                      1 & \ldots & 1 & 1 & \ldots & 1 \\
                      \psi_1 (x, t) & \cdots & \psi_N (x, t) & \psi_1^{-1} (x, t) & \ldots & \psi_N^{-1} (x, t) \\
                    \end{bmatrix}.
$$
According to \textbf{Theorem 2.1.}, the transfer matrix-function $W_A \left( x, \lambda \right) $ of the
dynamic system corresponding to matrix-function $S(x, t)$ has the form
\begin{equation}
    W_A \left( x, t; \lambda \right) = \prod^N_{i = 1} \left( \lambda ^2 - \alpha^2_{i, 0} \right)\{
    D_{kj} \left( x, t; \lambda \right) \} ^2 _{k, j = 1},
\end{equation}
where
\begin{center}
   $D_{11} \left( x, t; \lambda \right) = det\begin{bmatrix}
                                                            V_{N - 1} \left( 1 , \Omega \right) & V_{N - 1} \left( 1 , -\Omega \right) & 0 \\
                                                            V_{N} \left( X , \Omega \right) &  V_{N} \left( X^{-1} , -\Omega \right) & \Lambda_N \\
                                                               \end{bmatrix}
  \Delta^{-1} (x, t),$
  \end{center}
 \begin{center}
   $D_{12} \left( x, t; \lambda \right) = det\begin{bmatrix}
                                                            V_{N - 1} \left( 1 , \Omega \right) & V_{N - 1} \left( 1 , -\Omega \right) & \Lambda_{N - 1}  \\
                                                            V_{N} \left( 1 , \Omega \right) & V_{N} \left( 1 , -\Omega \right) & 0\\
                                                          \end{bmatrix}
    \Delta^{-1} (x, t),$
\end{center}

\begin{equation} \label{eq:DPoly1}
\end{equation}
 \begin{center}
   $D_{21} \left( x, t; \lambda \right) =  det\begin{bmatrix}
                                                            V_{N} \left( 1 , \Omega \right) & V_{N} \left( 1 , -\Omega \right) & 0 \\
                                                            V_{N - 1} \left( 1 , \Omega \right) & V_{N - 1} \left( 1 , -\Omega \right) & \Lambda_{N - 1} \\
                                                               \end{bmatrix}
  \Delta^{-1} (x, t),$
  \end{center}
 \begin{center}
   $D_{22} \left( x, t; \lambda \right) =  det\begin{bmatrix}
                                                            V_{N - 1} \left( 1 , \Omega \right) & V_{N - 1} \left( 1 , -\Omega \right) & 0 \\
                                                            V_{N - 1} \left( X , \Omega \right) &  V_{N - 1} \left( X^{-1} , -\Omega \right) & \Lambda_N \\
                                                          \end{bmatrix}
    \Delta^{-1} (x, t),$
\end{center} 
with
\begin{equation} \label{eq:DPoly2}
 \Delta (x, t) = det\begin{bmatrix}
V_{N - 1} \left( 1 , \Omega \right) & V_{N - 1} \left( 1 , -\Omega \right) \\
V_{N - 1} \left( X , \Omega \right) &  V_{N - 1} \left( X^{-1} , -\Omega \right) \\
\end{bmatrix}.
\end{equation}

$X = \left \{ \chi _k (x, t) \right \}_{k = 1}^N$ and operations on the sets $\Omega , \; X$ are assumed to be performed member-wise.\\

\begin{thm}
The following relation is true
\begin{equation} \label{eq:DWA}
    \frac{\partial}{\partial x}W_A \left( x, t; \lambda \right) = 2 \left( \lambda [j, W_A \left( x, t; \lambda
    \right)] + \Gamma _{2x} (2x, 0, t) j_1 W_A \left( x, t; \lambda
    \right) \right),
\end{equation}
where
\begin{center}
    $j = \begin{bmatrix}
            0 & 0 \\
            0 & 1 \\
          \end{bmatrix}, \;
    j_1 = \begin{bmatrix}
            0 & 1 \\
            1 & 0 \\
          \end{bmatrix};$
\end{center}
and $ [\cdot, \cdot]$ - is a commutator symbol defined by $ [M_1, M_2] = M_1 M_2 - M_2 M_1$.
\end{thm}
 
\begin{Rk}
Formula ~\eqref{eq:DWA} is an analogue of the formulas obtained in the series of papers \cite{sasa2, sasa3, sasa4, sasa5} where more general setup was considered.
\end{Rk}
\emph{Proof.} 
Differentiating~\eqref{eq:TMFA} with respect to $x$ we obtain
\begin{equation}
\frac{\partial}{\partial x}W_A \left( x, t; \lambda \right) = -\frac{\partial \Pi_2^T}{\partial x} S^{-1} (A - \lambda I_{2N})^{-1} \Pi_1 + \Pi_2^T S^{-1} \frac {\partial S}{\partial x} S^{-1} (A - \lambda I_{2N})^{-1} \Pi_1.
\end{equation}
It's easy to verify that
\begin{equation} \label{eq:JSB}
\begin{split}
\frac {\partial S}{\partial x} = 2 JSB, \\
\frac{\partial \Pi_2^T}{\partial x} = 2j\Pi_2^T B,
\end{split}
\end{equation}
where 
$$
J = \begin{bmatrix}
0 & 0 \\
0 & I_N \\
\end{bmatrix}.
$$
In our case differential equation for the matrix $S(x, t)$ is essentially different from the equation obtained in \cite{sasa2}-\cite{sasa5}, where $\partial S / \partial x$ is expressed in terms of $\Pi_1$ and $\Pi_2$.
Also from~\eqref{eq:identity} it follows that
\begin{equation} \label{eq:A}
A - SBS^{-1} = \Pi_1 \Pi_2^T S^{-1}. 
\end{equation} 
First consider the second term on the right hand side of ~\eqref{eq:DWA}. In view of ~\eqref{eq:JSB} we get
\begin{equation} \label{eq:233}
\begin{aligned}
& \Pi_2^T S^{-1} J S B S^{-1} (A - \lambda I_{2N})^{-1} \Pi_1\\
& = \Pi_2^T S^{-1} J (A - \lambda I_{2N} + \lambda I_{2N} - \Pi_1 \Pi_2^T S^{-1}) (A - \lambda I_{2N})^{-1} \Pi_1 \\
& = \Pi_2^T S^{-1} J \Pi_1 + \lambda \Pi_2^T S^{-1} J (A - \lambda E_{2N})^{-1} \Pi_1 - \Pi_2^T S^{-1} J \Pi_1 \underbrace{\Pi_2^T S^{-1} (A - \lambda I_{2N})^{-1} \Pi_1}_{I_2 - W_A ( x, t; \lambda )} \\
& = \lambda \Pi_2^T S^{-1} J (A - \lambda I_{2N})^{-1} \Pi_1 + \Pi_2^T S^{-1} J \Pi_1 W_A ( x, t; \lambda ) \\
& = \lambda \Pi_2^T S^{-1} (A - \lambda I_{2N})^{-1} \Pi_1 j + \Pi_2^T S^{-1} \Pi_1 j W_A ( x, t; \lambda ) \\
& = \lambda \left ( j - W_A ( x, t; \lambda ) \right ) + \Pi_2^T S^{-1} \Pi_1 j W_A ( x, t; \lambda ). \\
\end{aligned} 
\end{equation}
Substituting the relation $B S^{-1} = S^{-1} (A - \Pi_1 \Pi_2^T S^{-1})$, following from ~\eqref{eq:A}, into the first term on the right hand side of ~\eqref{eq:DWA} we obtain
\begin{equation} \label{eq:234}
\begin{aligned}
& j \Pi_2^T B S^{-1} (A - \lambda I_{2N})^{-1} \Pi_1\\
& = j \Pi_2^T S^{-1} ( A - \Pi_1 \Pi_2^T S^{-1} ) (A - \lambda I_{2N})^{-1} \Pi_1 \\
& = j \Pi_2^T S^{-1} (A - \lambda I_{2N} + \lambda I_{2N} - \Pi_1 \Pi_2^T S^{-1}) (A - \lambda I_{2N})^{-1} \Pi_1 \\
& = j \Pi_2^T S^{-1} \Pi_1 + \lambda j \underbrace{\Pi_2^T S^{-1} (A - \lambda I_{2N})^{-1} \Pi_1}_{I_2 - W_A ( x, t; \lambda )} 
- j \Pi_2^T S^{-1} \Pi_1 \Pi_2^T S^{-1} \underbrace{\Pi_2^T S^{-1} (A - \lambda I_{2N})^{-1} \Pi_1}_{I_2 - W_A ( x, t; \lambda )} \\
& = j \Pi_2^T S^{-1} \Pi_1 + \lambda j ( I_2 - W_A ( x, t; \lambda ) ) - j \Pi_2^T S^{-1} \Pi_1 ( I_2 - W_A ( x, t; \lambda ) ) \\
& = \lambda j ( I_2 - W_A ( x, t; \lambda ) ) + j \Pi_2^T S^{-1} \Pi_1 W_A ( x, t; \lambda ). \\
\end{aligned} 
\end{equation} 
Combining ~\eqref{eq:233} and ~\eqref{eq:234} yields
\begin{equation}
\frac{\partial}{\partial x}W_A ( x, t; \lambda ) = 2 \left ( \lambda [j, \; W_A ( x, t; \lambda )] + [\Pi_2^T S^{-1} \Pi_1, \; j] W_A ( x, t; \lambda ) \right ).
\end{equation}
Taking into account~\eqref{eq:Gamma}, expression $[\Pi_2^T S^{-1} \Pi_1, \; j]$ can be written as $\Gamma _{2x} (2x, 0, t) j_1$. This completes the proof. $\square$ \\
\begin{Rk}
Formula~\eqref{eq:DWA} establishes the connection between dynamic systems of the type~\eqref{eq:DynS} and solutions of NIDE.
\end{Rk}
Let's rewrite the quantities $\{ D_{k j} \left( x, t; \lambda \right) \} ^2 _{k, j = 1}$ (the elements of the matrix-polynomial in the representation of the transfer matrix-function ~\eqref{eq:sys213}) as polynomials with respect to $\lambda$ 
\begin{equation} \label{eq:DPoly}
\begin{split}
D_{1 1} \left( x, t; \lambda \right) = (-1)^N \sum_{i = 0}^N {(-1)^i a_i (x, t) \lambda ^{N - i}}, \; a_0 = 1; \\
D_{1 2} \left( x, t; \lambda \right) = \sum_{i = 0}^{N - 1} {b_i (x, t) \lambda ^{N - 1 - i}}; \\
D_{2 1} \left( x, t; \lambda \right) = (-1)^{N - 1} \sum_{i = 0}^{N - 1} {(-1)^i b_i (x, t) \lambda ^{N - 1 - i}};  \\
D_{2 2} \left( x, t; \lambda \right) = \sum_{i = 0}^N {a_i (x, t) \lambda ^{N - i}}, \; a_0 = 1. \\
\end{split}
\end{equation}
Substituting~\eqref{eq:DPoly} into~\eqref{eq:DWA}, one can establish the relationship between the coefficients $\left \{ a_i (x, t) \right \}_{i = 1}^N$ and $\left \{ b_i (x, t) \right \}_{i = 0}^{N - 1}$ of the polynomials. In this way
 we prove the following statement \\
\begin{thm}
  Let 
\begin{equation}
X(x, t) = col \left [ b_0 (x, t) \; \cdots \; b_{N - 1} (x, t) \; a_1 (x, t) \; \cdots \; a_N (x, t) \right ],
\end{equation}
where $\left \{ a_i (x, t) \right \}_{i = 1}^N$ and $\left \{ b_i (x, t) \right \}_{i = 0}^{N - 1}$ are the coefficients of the polynomials~\eqref{eq:DPoly}, then $X(x, t)$ satisfies the following Riccati-type system of differential equations
\begin{equation} \label{eq:Riccati}
\frac {\partial}{\partial x} X(x, t) = X(x, t) F X(x, t) - G X(x, t),
\end{equation}
where $F$ and $G$ are constant matrices

$$
F =  [\underbrace{0 \; 0 \; \cdots \; 0}_N \; \underbrace{2 \; 0 \; \cdots \; 0}_N  ],
$$ 
$$
\sbox0{$\begin{matrix}0&2&0&\cdots&0\\0&0&2&\cdots&0\\ \cdots&\cdots&\cdots&\cdots&\cdots\\0&0&0&\cdots&2\\0&0&0&\cdots&0\end{matrix}$}
G=\left[
\begin{array}{c|c}
\makebox[\wd0]{\large $0$}&\makebox[\wd0]{\large $0$}\\
\hline
  \vphantom{\usebox{0}}\makebox[\wd0]{\large $0$}&\usebox{0}
\end{array}
\right],
$$
and 
\begin{equation} \label{eq:b0}
b_0 (x, t) = -\frac {1}{2} \Gamma_{2x} (2x, 0, t).
\end {equation}
\end{thm}
\emph{Proof.} Let's rewrite the equations~\eqref{eq:Riccati} as
\begin{equation} \label{eq:Riccati2}
\begin{array}{c c}
b_0^{\prime} = 2 ( b_0 a_1 - b_1 ) &  a_1^{\prime} = 2 b_0^2 \\
b_1^{\prime} = 2 ( b_0 a_2 - b_2 ) &  a_1^{\prime} = 2 b_0 b_1 \\
\cdots \cdots \cdots \cdots \cdots &  \cdots \cdots \cdots \cdots \cdots \\
b_{N - 2}^{\prime} = 2 ( b_0 a_{N - 1} - b_{N - 1} ) & a_{N - 1}^{\prime} = 2 b_0 b_{N - 2} \\
b_{N - 1}^{\prime} = 2 b_0 a_{N - 1} &  a_N^{\prime} = 2 b_0 b_{N - 1} \\
\end{array}
\end{equation}
In~\eqref{eq:Riccati2} we omitted dependence on $x$ and $t$ and used 'prime' to designate the derivative with respect to $x$. Differentiating $\{ D_{k j} \left( x, t; \lambda \right) \} ^2 _{k, j = 1}$ and using~\eqref{eq:Riccati2} we get
$$
\begin{aligned}
D_{11}^{\prime} & = (-1)^N [-\lambda ^{N - 1} a_1^{\prime} + \lambda ^{N - 2} a_2^{\prime} + \cdots + (-1)^N a_N^{\prime}] \\
								& = 2 (-1)^N [-\lambda ^{N - 1} b_0^{2} + \lambda ^{N - 2} b_0 b_1 + \cdots + (-1)^{N - 1} b_0 b_{N - 1} ] = 2 b_0 D_{21}, \\
\end{aligned}
$$
$$
\begin{aligned}
D_{22}^{\prime} & = \lambda ^{N - 1} a_1^{\prime} + \lambda ^{N - 2} a_2^{\prime} + \cdots + a_N^{\prime}  \\
								& = 2 [\lambda ^{N - 1} b_0^{2} + \lambda ^{N - 2} b_0 b_1 + \cdots + b_0 b_{N - 1} ] = 2 b_0 D_{12}, \\
\end{aligned}
$$

$$
\begin{aligned}
D_{12}^{\prime} & =  \lambda ^{N - 1} b_0^{\prime} - \lambda ^{N - 2} b_1^{\prime} + \cdots + (-1)^{N - 1}b_{N - 1}^{\prime}  \\
								& = 2 \left [ \lambda ^{N - 1} (b_0 a_1 - b_1) + \lambda ^{N - 2} (b_0 a_2 - b_2) + \cdots + b_0 a_N \right ]  \\
								& = 2 ( b_0 \left [a_1 \lambda ^{N - 1} + a_2 \lambda ^{N - 2} + \cdots + a_N \right ] \\ 
							  & - [b_1 \lambda ^{N - 1} + b_2 \lambda ^{N - 2} + \cdots + \lambda b_{N - 1} ] ) \\
								& = 2 ( b_0 \left [ \lambda ^{N} + a_1 \lambda ^{N - 1} + a_2 \lambda ^{N - 2} + \cdots + a_N \right ] \\
								& - \lambda \left [ b_0 \lambda ^{N - 1} + b_1 \lambda ^{N - 2} + \cdots + b_{N - 1} \right ] )\\
								& = 2 ( b_0 D_{22} - \lambda D_{12} ),\\
\end{aligned}
$$
$$
\begin{aligned}
D_{21}^{\prime} & = (-1)^{N - 1} \left [ \lambda ^{N - 1} b_0^{\prime} - \lambda ^{N - 2} b_1^{\prime} + \cdots + (-1)^{N - 1}b_{N - 1}^{\prime} \right ] \\
								& = 2 (-1)^{N - 1} \left [ \lambda ^{N - 1} (b_0 a_1 - b_1) - \lambda ^{N - 2} (b_0 a_2 - b_2) + \cdots + (-1)^{N - 1} b_0 a_N \right ]  \\
								& = 2 (-1)^{N - 1} ( b_0 \left [a_1 \lambda ^{N - 1} - a_2 \lambda ^{N - 2} + \cdots + (-1)^{N - 1} a_N \right ] \\ 
							  & - \left [ b_1 \lambda ^{N - 1} - b_2 \lambda ^{N - 2} + \cdots + (-1)^{N - 2} \lambda b_{N - 1} \right ] ) \\
								& = 2 (-1)^{N - 1} ( b_0 \left [- \lambda ^{N} + a_1 \lambda ^{N - 1} - a_2 \lambda ^{N - 2} + \cdots + (-1)^{N - 1} a_N \right ] \\
								& - \lambda \left [ -b_0 \lambda ^{N - 1} + b_1 \lambda ^{N - 2} - \cdots + (-1)^{N - 2} b_{N - 1} \right ] )\\
								& = 2 ( b_0 D_{11} + \lambda D_{21} ).\\
\end{aligned}
$$
This is equivalent to~\eqref{eq:DWA}. 
By comparing~\eqref{eq:Gamma1} -~\eqref{eq:Gamma3}, ~\eqref{eq:DPoly1}, ~\eqref{eq:DPoly2} and ~\eqref{eq:DPoly}, it's easy to see that~\eqref{eq:b0} is valid. $\square$
\begin{Rk}
Formula \eqref{eq:b0} establishes the connection between the coefficients of the polynomials $\{ D_{k j} \left( x, t; \lambda \right) \} ^2 _{k, j = 1}$ and solutions of NIDE.
\end{Rk}
Potentials corresponding to the solutions of NIDE of the type~\eqref{eq:Gamma} in \cite{sasa7} are called Pseudo Exponential (PE) so our solutions of NIDE can be considered as an analogue of PE potentials. In further considerations we'll be using the notation $PE(N)$ to reflect the fact that the potential is parametrized by $2N$ parameters according to the Theorems 3.3 and 3.4. In \cite{sasa7} it was given a characterization of PE potentials in terms of their Taylor coefficients and reflection coefficient. We re-formulate and prove this result in the context of our case. 
\begin{thm}
Let $b_0(x, t) \; (\Gamma_{2x}(2x, 0, t))$  at some point $t = t_0$ be a $PE(N)$ potential, meromorphic on $\mathbb{R} \times [0, \infty]$ and analytic at $(x_0, t_0)$. Then it is uniquely defined by \\
$b_0(x_0, t_0), b_0^{\prime}(x_0, t_0),  \cdots, b_0^{(2N - 1)}(x_0, t_0)$ (the derivatives are taken with respect to $x$).
\end{thm}
\emph{Proof.} First, let's fix $t: t = t_0$. The problem then reduces to the reconstruction of the $PE(N)$ potential, in other words, to build two sets of parameters
 $\Omega = \left \{ \omega_k \right \}_{k = 1}^N; \; \A_0 = \left \{ \alpha_{k, 0} \right \}_{k = 1}^N$ given its first $2N - 1$ derivatives at some point of analyticity $x_0$. The procedure is based on the relations~\eqref{eq:Riccati2}. For the convenience we perform the following transformation of the variable $x$: $x \rightarrow 2 x$. By differentiating equations~\eqref{eq:Riccati2} $2 N - 1$ times at $x_0$ we arrive at the system of $2 N - 1$ linear equations $C X = Y$ with respect to the quantities $X = \left \{ x_k \right \}_{k = 1}^{2 N - 1} \equiv \left \{ a_1 \; \cdots \; a_N \; b_1 \; \cdots \; b_{N - 1} \right \}$. Elements of the matrix $C = \left \{ c_{i, j} \right \}_{i, j = 1}^{2N - 1}$ and vector $Y = col [y_1 \; y_2 \; \cdots \; y_{2N - 1}]$ are calculated as follows
   
\[c_{i, j} = 
  \begin{cases}
0; & j > i; \: j < N; \\
c_{i - 1, j}^{\prime} + b_0 \: c_{i - 1, j + N -1}; & i \geq j, \: j < N; \\
(-1)^{i}; & j - i = N, \: N < j \leq 2 N - 1; \\
0; & j - i = N - 1, \: N < j \leq 2 N - 1; \\
c_{i - 1, j}^{\prime} + b_0 \: c_{i - 1, j - N} - c_{i - 1, j - 1}; & j - i \leq N - 2, \: N < j \leq 2 N - 1; \\
\end{cases}
\]
$$
y_i = b_0^{(i)} - b_0^{(i - 2)} b_0^2 - y_{i - 1}^{\prime}; \: 1 \leq i \leq 2N - 1.
$$
If $det|C| \neq 0$ then this system has a unique solution. Then according to~\eqref{eq:DPoly} we construct the polynomials $\{ D_{k j} \left( x_0, \lambda \right) \} ^2 _{k, j = 1}$. Applying \textbf{Theorem 2.3} to this special case, we reduce the problem to \textbf{IIP Problem} that has a unique solution given by the following procedure:
\begin{itemize}
\item Find the roots $ \left \{ \tilde{\omega}_k \right \}_{k = 1}^N;$ of the polynomial
$$
D(x_0, \lambda) = D_{11}(x_0, \lambda) D_{22}(x_0, \lambda) - D_{12}(x_0, \lambda) D_{21}(x_0, \lambda)
$$
and set $\omega_k = \tilde{\omega}_k; \: 1 \leq k \leq N$
\item Using relations~\eqref{eq:sys235}-~\eqref{eq:sys237} compute the ratios 
$$
R = \frac {D_{12} (\omega_k)}{D_{11} (\omega_k } = \frac {D_{22} (\omega_k)}{D_{21} (\omega_k },
$$
which can be considered as a system of linear equations with respect to the elementary symmetric forms $\sigma (A_0)$ where $A_0 = \left \{\alpha _k (0) \right \}_{k = 1}^N$;
\item By solving this system and then finding the roots of the polynomial
$$
P(\sigma (A_0), \lambda) = \sum_{k = 0}^N {\sigma_k (A_0) \lambda ^{N - k}}; \: i = 1
$$
one recovers the set $A_0$. $\square$
\end{itemize}
The above result can be re-formulated in terms of the inverse problem for the solution of NIDE.\\
\textbf{Inverse NIDE problem}. \emph{Let $\xi(x, t)$ be a solution of NIDE such that at some point $t = t_0 \; \xi(x, t_0) \in PE(N)$. Given the derivatives\\
 $\xi(x_0, t_0), \xi^{\prime}(x_0, t_0),  \cdots, \xi^{(2N - 1)}(x_0, t_0)$ at some point $x_0$ of analyticity of $\xi(x, t)$, restore the solution $\xi(x, t)$ on $\mathbb{R} \times [0, \infty]$}.\\
\begin{Rk}
 \textbf{Theorem 3.7} solves the \textbf{Inverse NIDE problem}.
\end{Rk}
\begin{Rk}
The form of the solution of SHG equation $\phi(x, t)$ in spatial variable $x$ differs from the function $\Gamma_{2x}(2x, 0, t)$ by one extra derivative and constant multiplier, which suggests a slight modification to the procedure described above: it requires the derivatives of order $1, 2, \ldots, 2N$ for the solution of the \textbf{Inverse NIDE problem}. Without loss of generality, in further considerations we'll be referring to the \textbf{Inverse NIDE problem} as applied to the function $\xi(x, t) = \Gamma_{2x}(2x, 0, t)$ 
\end{Rk}
To illustrate the methodology consider function $\xi(x, t)$ when $N = 1$. 

\begin{Ee}
Let $t_0 = 0$, then
\begin{equation}
\xi(x, 0) \equiv \xi(x) = \omega \csch (2 \omega x - \ln |\frac {\omega - \alpha_0}{\omega + \alpha_0}|).
\end{equation} 
It's easy to verify that the equations 
\begin{equation}
\xi(x)^{\prime} = 2 \xi (x) a_1, \; a_1(x)^{\prime} = 2 \xi(x)^2
\end{equation} 
derived from~\eqref{eq:Riccati}, are satisfied with $a_1(x) = -\omega \coth (2 \omega x - \ln |\frac {\omega - \alpha_0}{\omega + \alpha_0}|)$. Then we construct the polynomials $\{ D_{k j} \left( x, \lambda \right) \} ^2 _{k, j = 1}$
\begin{equation}
\begin{aligned}
D_{11}(x, \lambda) = & - \lambda + a_1(x); D_{12}(x, \lambda) = \xi(x); \\
D_{22}(x, \lambda) = & \lambda + a_1(x);  D_{21}(x, \lambda) = \xi(x); \\
\end{aligned}
\end{equation}
and calculate the roots $\lambda_{1, 2}$ of the polynomial
\begin{equation}
D(x, \lambda) = (- \lambda + a_1(x) ) ( \lambda + a_1(x) ) - \xi(x)^2 = -\lambda^2 + a_1(x)^2 - \xi(x)^2.
\end{equation}
After elementary calculations we find that $\lambda_{1, 2} = \pm \omega$. By computing the ratios
$$
R = \frac{\xi(x)}{-\omega + a_1(x)} = \frac{\omega + a_1(x)}{\xi(x)},
$$
we obtain the quantity $\ln R = 2\omega x - \ln \frac{\omega - \alpha_0}{\omega + \alpha_0}$ from which it's easy to calculate $\alpha_0$ as
$$
\alpha_0 = \omega \frac{1 - R_1}{1+ R_1}; \; R_1 = \exp (2\omega x ) / R.
$$
It's also easy to see that $R_1$ doesn't depend on $x$.
\end{Ee}
\begin{Ee}
Let $N = 2$. Given the point $x_0$ and numbers $\xi(x_0), \xi^{\prime}(x_0), \xi^{\prime \prime}(x_0), \xi^{\prime \prime \prime}(x_0)$ we show how to construct the coefficients $a_1, a_2, b_1$ of the polynomials $\{ D_{k j} \left( x, \lambda \right) \} ^2 _{k, j = 1}$. Matrix $C$ and vector $Y$ have the following representation
\begin{equation}
C =
\begin{bmatrix}
		\xi_0 & 0 & -1 \\
		\xi_0^{\prime} & -\xi_0 & 0 \\
		\xi_0^{\prime \prime} & -\xi_0^{\prime} & -\xi_0^2 \\
\end{bmatrix}; \;
Y = col [\xi_0^{\prime}, \: \xi_0^{\prime \prime} - \xi_0^3, \: \xi_0^{\prime \prime \prime} - 4 \xi_0^{\prime} \xi_0^2 ].
\end{equation}
The solution $X = col [a_1, \: a_2, \: b_1]$ of the system $C X = Y$ is
\begin{equation} 
\begin{aligned}
a_1 = & \frac{\xi_0^{\prime} \xi_0^{\prime \prime} - \xi_0 \xi_0^{\prime \prime \prime} + 4 \xi_0^{\prime} \xi_0^3}{\xi_0^4 + {\xi_0^{\prime}}^2 - \xi_0 \xi_0^{\prime \prime}}; \\
a_2 = & \frac{4 {\xi_0^{\prime}}^2 \xi_0^2 + {\xi_0^{\prime \prime}}^2 - \xi_0^{\prime} \xi_0^{\prime \prime \prime} - 2 \xi_0^3 \xi_0^{\prime \prime} + \xi_0^6}{\xi_0^4 + {\xi_0^{\prime}}^2 - \xi_0 \xi_0^{\prime \prime}}; \\
b_1 = & \frac{4 \xi_0^4 \xi_0^{\prime} - \xi_0^2 \xi_0^{\prime \prime \prime} - {\xi_0^{\prime}}^3 + \xi_0 \xi_0^{\prime} \xi_0^{\prime \prime}}{\xi_0^4 + {\xi_0^{\prime}}^2 - \xi_0 \xi_0^{\prime \prime}} \\
\end{aligned}
\end{equation}
\end{Ee}
It's interesting to note that there is a connection between the solutions of the considered NIDE and other non-linear differential equations. For example, Miura transformation
\begin{equation}\label{eq:Miura}
M[f(x)] = f(x)^2 \pm \frac{\mathrm d f(x)}{\mathrm d x}
\end{equation}
 converts solutions of MKdV equation into the solutions of Korteweg - deVries (KdV) equation 
\begin{equation}\label{eq:KdV}
\frac {\partial u (x, t)}{\partial t} = - \frac{1}{4} \frac {\partial ^3 u (x, t)}{\partial x ^3} +
\frac{3}{2} | u (x, t) | \frac {\partial u (x, t)}{\partial x},
\end{equation}
and solutions of NSE again into the solutions of NSE with opposite sign by the non-linear term. The corresponding image $M[\xi(x, t)]$ can be represented in the standard form 
\begin{equation}
P(x, t) = -2 \frac{\partial ^2 \ln \left (\delta (x, t) \right )}{\partial x^2}
\end{equation}
where
\[ \delta (x, t) =
  \begin{cases}
    \delta_2 (x, t)       & \; \text{when choosing "+" in~\eqref{eq:Miura} }, \\
    \delta_1 (x, t)       & \; \text{when choosing "-" in~\eqref{eq:Miura} }; \\
  \end{cases}
\]
and $\delta_{1, 2} (x, t)$ are defined by~\eqref{eq:Delta1},~\eqref{eq:Delta2}. If $\delta (x, t) \neq 0; \; \forall (x, t) \in (-\infty, \infty )$ then $P(x, t)$ - is the N-soliton solution of the corresponding non-linear equation. We illustrate the above assertions by simple examples.
\begin{Ee}
Consider the solution of MKdV for the case $N = 1$
\begin{equation}
\psi (x, t) = 2 \omega \csch \left ( 2 \chi (x, t) \right)
\end{equation}
where $\chi (x, t) =  \omega x - \omega ^3 t - 1/2 \ln |(\omega - \alpha_0 ) / (\omega + \alpha_0 )|$. It's easy to check by direct computation that the function
\begin{equation}
P_1(x, t) = \psi (x, t) ^2 + \frac{\partial \psi (x, t)}{\partial x} = -2 \omega ^2 \sech^2 \left ( \chi (x, t) \right ) 
\end{equation} 
satisfies KdV equation~\eqref{eq:KdV} with $\delta (x, t) = \cosh \left (\chi (x, t) \right )$. Analogously, function
\begin{equation}
P_2(x, t) = \psi (x, t) ^2 - \frac{\partial \psi (x, t)}{\partial x} = -2 \omega ^2 \csch^2 \left ( \chi (x, t) \right ) 
\end{equation} 
satisfies KdV equation~\eqref{eq:KdV} with $\delta (x, t) = \sinh \left (\chi (x, t) \right )$.
\end{Ee}
In \textbf{Example 3.13} $P_1(x, t)$ represents a classical 1-soliton solution of KdV equation. As opposed to $P_1(x, t)$, function $P_2(x, t)$ is singular on $\mathbb{R} \times [0, \infty]$ and doesn't belong to N-soliton family, but because of the similar nature we'll refer to the Miura-transformed PE(N)-functions as \emph{soliton-like} (SL(N)) solutions of NIDE.  
Combining results obtained in \textbf{Theorem 3.7} and properties of Miura-transformed PE(N)-functions, we can solve an inverse problem for the SL(N) solutions of NIDE.\\
\begin{thm}
 Let $q(x, t)$ be SL(N) solution of NIDE, meromorphic on $\mathbb{R} \times [0, \infty]$ and analytic at $(x_0, t_0)$. Then it is uniquely defined by \\
$q(x_0, t_0), q^{\prime}(x_0, t_0),  \cdots, q^{(2N - 1)}(x_0, t_0)$ (the derivatives are taken with respect to $x$).
\end{thm}
\emph{Proof}. First, as in \textbf{Theorem 3.7}, let's fix $t : t = t_0$. Using relations~\eqref{eq:Riccati2} and Miura transformation~\eqref{eq:Miura} by consecutive differentiation of the system~\eqref{eq:Riccati2} we arrive at the system of equations $\tilde{C} \tilde{X} = \tilde{Y}$ with respect to the quantities $X = \left \{ x_k \right \}_{k = 1}^{2 N - 1} \equiv \left \{ a_1 \; \cdots \; a_N \; b_0 \; \cdots \; b_{N - 1} \right \}$. Matrix $\tilde{C}$ of the size $2 N \times N$ is represented in block form as
$$ \tilde{C} = 
\begin{bmatrix}
       \tilde{C}_1  \\
       \tilde{C}_2  
\end{bmatrix},
$$
where the elements of the matrices $\tilde{C}_i; \; i = 1, 2$ of the size $N \times N$ each, are 
computed as follows. For the matrix $\tilde{C}_1$ we have
\[c_{i, j} = 
  \begin{cases}
0 & j > i + 1; \: j \leq N; \\
(-1)^{j - 1} & j = i + 1; \: j \leq N; \\
(-1)^{i - 1} & j = i, \: j \leq N; \\
(-1)^{i - 1}c_{i + 1, j + 1} & 1 \leq i, j \leq N;\\
(-1)^{j - 1} \vartheta_{N - j} (b_{N - j - 1} + a_{N - j}) & i = N, \: 1 \leq j \leq N; \\
\end{cases}
\]
where quantities $\vartheta_{N - j}$ are constructed as 
$$
\vartheta_{-1} = 1; \: \vartheta_{0} = b_0; \: \vartheta_{1} = q_0 = b_0^2 \pm b^{\prime}_0; \: \vartheta_{2} = -q^{\prime}_0; \: \vartheta_{3} = q^{\prime \prime}_0 - q_0^2;
$$
\begin{equation}  \label{eq:vartheta}
\end{equation}
$$
\vartheta_{i + 1} = -\vartheta^{\prime}_{i} - \sum_{j = 1}^{i - 1}{\vartheta_{i - j}}\vartheta_{j} \: (i = 2, 3, \ldots).
$$
Corresponding vector of unknowns $\tilde{X}_1$ is organized in the following way 
$$
x_i = b_{i - 1} + a_i; \: 1 \leq i \leq N,
$$
and elements of the vector $\tilde{Y}_1$ are computed as
$$
y_i = (-1)^{i - 1}\vartheta_{i}; \: i = 1, 2, \cdots, N.
$$
For the matrix $\tilde{C}_2$ we have
\[c_{i, j} = 
  \begin{cases}
(-1)^{i - 1}c_{i + 1, j + 1}, & 1 \leq i, j \leq N;\\
\vartheta_{N - j + 1} & i = 1, \: 1 \leq j \leq N; \\
\end{cases}
\]
Corresponding vector of unknowns $\tilde{X}_2$ is  
$$
x_i = b_{i - 1} + a_i; \: 1 \leq i \leq N
$$
and elements of the vector $\tilde{Y}_2$ are computed as
$$
y_i = \vartheta_{N + i}; \: i = 1, 2, \ldots, N
$$
The system is solved in four simple steps:
\begin{itemize}
\item If $det[\tilde{C}_2] \neq 0$ then the system $\tilde{C}_2 \tilde{X}_2 = \tilde{Y}_2$ has a unique solution. Solving this system we find the quantities $d_i = b_{i - 1} + a_i; \: 1 \leq i \leq N$;
\item Substituting $d_i; \: 1 \leq i \leq N$ into the last equation of the system $\tilde{C}_1 \tilde{X}_1 = \tilde{Y}_1$, we compute $b_0$;
\item Propagating backwards from $(N - 1)-th$ to the first equation in the system $\tilde{C}_1 \tilde{X}_1 = \tilde{Y}_1$, we calculate $b_i; 1 \leq i \leq N - 1$;
\item Compute $a_i = d_i - b_{i - 1}; 1 \leq i \leq N$.
\end{itemize}
The rest of the procedure is the same as in \textbf{Theorem 3.7}. $\square$\\
We illustrate the calculation steps by an example.
\begin{Ee}
Let $N = 2$ and given the quantities $q_0, q^{\prime}_0, q^{\prime \prime}_0, q^{\prime \prime \prime}_0$. 
Then
\begin{equation}
\begin{cases}
    q_0 = b_0\left(a_1 + b_0\right) - b_1;\\
    q^{\prime}_0 = q_0\left(a_1 + b_0\right) - b_0\left(a_2 + b_1\right);\\
    q^{\prime \prime}_0 = q^{\prime}_0\left(a_1 + b_0\right) - q_0\left(a_2 + b_1\right) +
    q^2_0;\\
    q^{\prime \prime \prime}_0 = \left(q^{\prime \prime}_0 - q^2_0\right)\left(a_1 + b_0\right) - q^{\prime}_0\left(a_2 +
    b_1\right) + 4 q_0 q^{\prime}_0.
\end{cases}
\end{equation}
From the last two equations we obtain
\begin{center}
    $d_1 = a_1 + b_0 = \frac{-q^{\prime}_0\left(q^{\prime \prime}_0 - q^2_0\right) + q_0\left(q^{\prime \prime \prime}_0 - 4 q_0 q^{\prime}_0\right)}{-q^{\prime 2}_0 + q^{\prime \prime}_0 q_0 - q^3_0}$;
\end{center}
\begin{center}
    $d_2 = a_2 + b_1 = \frac{q^{\prime}_0\left(q^{\prime \prime \prime}_0 - 4 q_0 q^{\prime}_0\right) - \left(q^{\prime \prime}_0 - q^2_0\right)^2}{-q^{\prime 2}_0 + q^{\prime \prime}_0q_0 - q^3_0}$.
\end{center}
And from the first two equations we have
\begin{center}
    $b_0 = \frac{q^{\prime}_0 - q_0 d_1}{d_2}$;
\end{center}
\begin{center}
    $a_1 = d_1 - b_0$;
\end{center}
\begin{center}
    $b_1 = d_1 b_0 - q_0$;
\end{center}
\begin{center}
    $a_2 = d_2 - d_1 b_0 + q_0$.
\end{center}
\end{Ee}
A thorough analysis of reflectionless (RL) potentials in Sturm-Liouville problem is given in \cite{mar}. In particular, it was considered a closure of the sets of RL potentials in the topology of uniform convergence of the functions on every compact of the real axis. These results are important in the problems of approximation of the functions by RL potentials. The criteria are given in terms of functions $\vartheta_{j}(x), \; j = -1, 0, 1, \ldots $ defined by the relations~\eqref{eq:vartheta}. Let $\mathcal{B}(-\mu^2) (\mu \geq 0)$ represent a set of RL potentials for which the spectrum of the corresponding operators lies to the right of the point $-\mu^2$ and $\mathcal{B}$ represents the set of all RL potentials i.e. $\mathcal{B} = \bigcup_{\mu \geq 0}{\mathcal{B}(-\mu^2)}.$ The following assertion is true (the proof is beyond the scope of this paper and can be found in \cite{mar}).\\
\begin{thm}
 For the real function $q_0(x)$ to belong to the set $\mathcal{B}$ it is necessary and sufficient that it is infinitely smooth at point $x$ and there exists a number $R < \infty$ such that defined by relations~\eqref{eq:vartheta} functions $\vartheta_{j}(x), \; j = -1, 0, 1, \ldots $ satisfy the inequalities 
\begin{equation} \label{eq:vartheta2}
|\vartheta_{j}(x)| \leq (2 R)^j R.
\end{equation}
\end{thm}
\begin{Cy}If for some function $\tilde{q}_0(x)$ conditions~\eqref{eq:vartheta2} are satisfied then it can be approximated by RL potentials with given accuracy.
\end{Cy}
\textbf{Theorem 3.8.} extends the results of \cite{mar} for the case of SL(N) solutions of NIDE. 

\subsection{Dynamics of the singularities of the PE(N) and SL(N) solutions of NIDE.}
As mentioned above, PE(N) and SL(N) solutions of NIDE can have singularities.
\begin{Dn}
Point $(x_0, t_0)$ on the plain $(x, t), \; -\infty < x, t < \infty$ is called a singularity point if $|\xi(x, t)| \to \infty $ when $x \to x_0$ and $t \to t_0$ where $\xi(x, t)$ is the solution of NIDE.
\end{Dn}
A set of singularity points on $(x, t)$ - plain is called a \emph{singularity line}. Dependence of the singularity point on the parameter $t$ forms a \emph{singularity line}. In the next section we investigate the dynamics of the \emph{singularity lines} of the PE(N) and SL(N) solutions of NIDE.

In \cite{lev2} the following assertion is proved
\begin{thm}
If $v_0(z) \in \mathcal{P}$, where $v_0(z) = v(0, z)$ and $v(t, z)$ is Weyl-Titchmarsh function of the system (2.4), (2.5), then the solution $\xi(x, t)$ of NIDE is regular in the region $(x, t) \geq 0$.
\end{thm}

It follows from the relations~\eqref{eq:sys325} and ~\eqref{eq:SH} that \emph{singularity lines} of the solution $\xi(x, t)$ of NIDE satisfy the equations
\begin{equation} \label{eq:slines}
\delta_j(x, t) = 0, \; j = 1, 2;
\end{equation} 
so the investigation of the dynamics of the singularities is equivalent to the study of the properties of the solutions of the system~\eqref{eq:slines}. Material of this section is based on the results obtained in \cite{dis}. Some of the proofs will be omitted here due to the simplicity.\\
First, we look at the asymptotics of the \emph{singularity lines} when $t \to \pm \infty$. It's easy to verify that the following assertion is true

\begin{An}
Let $x$ and $t$ be such that 
\begin{equation}
0 < \delta < |\chi_j(x, t) \pm \frac{1}{2} \ln{A_j}| < \epsilon, \; A_j = \prod_{i = 1}^{j - 1}{\frac{\omega_j + \omega_i}{\omega_j - \omega_i}} \prod_{i = j + 1}^{N}{\frac{\omega_i - \omega_j}{\omega_i + \omega_j}},
\end{equation}
then for the solution $\phi(x, t)$ of sinh-Gordon equation when $t \to \pm \infty$ the following representation is valid
\begin{equation} \label{eq:asg}
\phi(x, t) = 2 (-1)^{N + j} \ln{|\tanh{(\chi_j(x, t) \pm \frac{1}{2} \ln{A_j} } + O(1)|}, \; j = 1, 2, \ldots, N.
\end{equation}
Here $\chi_j(x, t) = \omega_j x + \frac{1}{\omega_j} t - \frac{1}{2} \ln{C_j}$ and $C_j = \prod _{i = 1}^N |\frac {\omega_j - \alpha_{i, 0}} {\omega_j + \alpha_{i, 0}}|$.
\end{An}
 From \textbf{Assertion 3.17} immediately follow the corollaries

\begin{Cy} For the sufficiently large values of $|t|$ and $t < 0$ function $\phi(x, t)$ has singularities in the region $|\chi_j(x, t) - \frac{1}{2} \ln{A_j}| < \epsilon$. 
\end{Cy}
\begin{Cy} For the sufficiently large values of $t$ function $\phi(x, t)$ has singularities in the region $|\chi_j(x, t) + \frac{1}{2} \ln{A_j}| < \epsilon$.
\end{Cy}
Analogous result takes place for the solutions $\psi(x, t)$ and $\rho(x, t)$ of the equations (3.2) and (3.3).
\begin{An}
Let $x$ and $t$ be such that 
\begin{equation}
0 < \delta < |\Re{\chi_j(x, t)} \pm \frac{1}{2} \ln{|A_j|}| < \epsilon, \; A_j = (-1)^{N - 1}\prod_{i = 1}^{j - 1}{\frac{\bar{\omega}_j + \omega_i}{\omega_j - \omega_i}} \prod_{i = j + 1}^{N}{\frac{\omega_i - \omega_j}{\bar{\omega}_i + \omega_j}},
\end{equation}
then the solution $\xi(x, t)$ when $t \to \pm \infty$ can be represented as
\begin{equation} \label{eq:akdv}
\xi(x, t) = \frac{2 (-1)^{N + j} \Re{\omega_j} \exp{\imath (\Im{\chi_j(x, t)} - \arg{A_j)}}} {\sinh{(\Re{\chi_j(x, t)} \pm \ln{|A_j|})}} + O(1), \; j = 1, 2, \ldots, N.
\end{equation}
Here $\xi(x, t) \equiv \psi(x, t), \; \chi_j(x, t) = 2 (\omega_j x - \omega_j^3 t - \frac{1}{2} \ln{C_j})$ in case of mKdV equation and $\xi(x, t) \equiv \rho(x, t), \; \chi_j(x, t) = 2 (\omega_j x + \imath \omega_j^2 t - \frac{1}{2} \ln{C_j})$ in case of NSE equation.
\end{An}
From \textbf{Assertion 3.20} immediately follow the corollaries
\begin{Cy} For the sufficiently large values of $|t|$ and $t < 0$ function $\xi(x, t)$ has singularities in the region $|\Re{\chi_j(x, t)} - \ln{A_j}| < \epsilon$. 
\end{Cy}
\begin{Cy}For the sufficiently large values of $t$ function $\xi(x, t)$ has singularities in the region $|\Re{\chi_j(x, t)} + \ln{A_j}| < \epsilon$. 
\end{Cy}
From asymptotic formulas~\eqref{eq:asg} and~\eqref{eq:akdv} it follows that if $x$ is considered as spacial and $t$ - as temporal variables then the solutions $\phi(x, t)$, $\psi(x, t)$, $\rho(x, t)$ when $t \to \pm \infty$ are represented as a complex of $N$ elementary singular waves. These waves interact, and after the interaction they preserve their shapes. The only change they suffer is the phase shift $\Delta_j = \ln|A_j|$. This behavior is quite similar to the behavior of the classical soliton solutions. Presence of singularities and soliton-like nature of their interaction suggests that the solutions $\phi(x, t)$, $\psi(x, t)$, $\rho(x, t)$ can be treated in terms of particles interacting by their surrounding field and corresponding \emph{singularity lines} can be identified as \emph{world lines} of the particles. Consider some simple examples.
\begin{Ee} In case $N = 1$ and $\omega = \bar{\omega}, \; \alpha_{0} = \bar{\alpha}_{0}$ we have one \emph{singularity line} that satisfies the equation
\begin{equation} \label{eq:sl1}
\omega x + \Theta (\omega) t -\frac {1}{2} \ln (|\frac {\omega - \alpha_{0}} {\omega + \alpha_{0}}|) = 0.
\end{equation} 
Equation~\eqref{eq:sl1} represents a straight line that corresponds to the world line of "free" particle propagating with velocity $v = \Theta (\omega) / \omega$.
\end{Ee}
In \cite{dis} the following assertion has been proved
\begin{An}$\Res{(\partial \phi(x, t) / \partial x)} = \pm 1$, $\Res{(\psi(x, t))} = \pm 1$ and $\Res{(\rho(x, t))} = \pm 1$
\end{An}
In "particle language" this means that there are two types of particles (corresponding to the sign of the residue). The following example demonstrates the interaction between particles with different combinations of the types. We consider solutions $\phi(x, t)$ of SHG equation (conceptually, the dynamics of \emph{singularity lines} in case of mKdV and NSE equations is the same).
\begin{Ee}
When $N = 2$ we consider three cases 
\begin{enumerate}
\item $\omega_i = \bar{\omega}_i, \; \alpha_{i, 0} = \bar{\alpha}_{i, 0}, \; C_i > 0, \; i = 1, 2$;
\item $\omega_i = \bar{\omega}_i, \; \alpha_{i, 0} = \bar{\alpha}_{i, 0}, \; i = 1, 2; \; C_1 < 0, \; C_2 > 0$;
\item $\omega_2 = \bar{\omega}_1, \; \alpha_{i, 0} = \bar{\alpha}_{i, 0}, \; i = 1, 2$;
\end{enumerate}
where $C_j = \frac {(\omega_j - \alpha_{1, 0})(\omega_j - \alpha_{2, 0})} {(\omega_j + \alpha_{1, 0})(\omega_j + \alpha_{2, 0})}, \; j = 1, 2$. In all the cases there are two \emph{singularity lines}. In case 1. solution $\phi(x, t)$ has the form
\begin{equation} \label{eq:shgsh}
\phi(x, t) = 2 \ln \left | \frac{(\omega_1 - \omega_2) \sinh(\eta_1(x, t)) - (\omega_1 + \omega_2) \sinh(\eta_2(x, t))}{(\omega_1 - \omega_2) \sinh(\eta_1(x, t)) + (\omega_1 + \omega_2) \sinh(\eta_2(x, t))} \right |,
\end{equation} 
where $\eta_1(x, t) = \chi_1(x, t) + \chi_2(x, t), \; \eta_2(x, t) = \chi_2(x, t) - \chi_1(x, t)$. In this case \emph{singularity lines} satisfy the equations
\begin{equation} \label{eq:sl21}
X_{1, 2} = \pm \sinh^{-1}\left ( \frac{\omega_1 + \omega_2}{\omega_2 - \omega_1} \sinh(Y)\right ),
\end{equation} 
where 
$$
X = (\omega_1 + \omega_2) x + (\Theta (\omega_1) + \Theta (\omega_2)) t -\frac{1}{2} \ln|C_1 C_2|;
$$
$$
Y = (\omega_2 - \omega_1) x + (\Theta (\omega_2) - \Theta (\omega_1)) t -\frac{1}{2} \ln|\frac{C_2}{ C_1}|.
$$
In case 2. solution $\phi(x, t)$ has the form
\begin{equation} \label{eq:shgch}
\phi(x, t) = 2 \ln \left | \frac{(\omega_1 - \omega_2) \cosh(\eta_1(x, t)) - (\omega_1 + \omega_2) \cosh(\eta_2(x, t))}{(\omega_1 - \omega_2) \cosh(\eta_1(x, t)) + (\omega_1 + \omega_2) \cosh(\eta_2(x, t))} \right |,
\end{equation} 
and \emph{singularity lines} satisfy the equations
\begin{equation} \label{eq:sl22}
X_{1, 2} = \pm \cosh^{-1}\left ( \frac{\omega_1 + \omega_2}{\omega_2 - \omega_1} \cosh(Y)\right ).
\end{equation}
In case 3. solution $\phi(x, t)$ is represented as
\begin{equation} \label{eq:shgcmplx}
\phi(x, t) = 2 \ln \left | \frac{\Im \omega_1 \sinh(\zeta_1(x, t) + \Re\omega_1 \sin(\zeta_2(x, t))}{\Im \omega_1 \sinh(\zeta_1(x, t) - \Im \omega_1 \sin(\zeta_2(x, t))} \right |.
\end{equation} 
Here $\zeta_1(x, t) = 2 \Re \chi(x, t), \; \zeta_2(x, t) = 2 \Im \chi(x, t)$ and $\chi(x, t) = \omega_1 x + \frac{1}{\omega_1} t - \frac{1}{2}\ln |C_1|$. Corresponding \emph{singularity lines} satisfy the equations
\begin{equation} \label{eq:sl23}
X_{1, 2} = \pm \sinh^{-1}\left ( \frac{\Re{\omega_1}}{\Im{\omega_1}} \sin(Y)\right ),
\end{equation} 
where
$
X = \zeta_1(x, t), \; Y = \zeta_2(x, t).
$
\end{Ee}
\emph{Singularity lines} corresponding to those three cases are depicted on the figures 1, 2 and 3 respectively in \textbf{Appendix}.\\
Case 1. presents the interaction of the particles of different types. When the values of $|t|$ are large and $t < 0$ the lines are close to the straight lines corresponding to the asymptotic solutions~\eqref{eq:asg} and~\eqref{eq:akdv} when $t \to -\infty$. Then the lines become closer and intersect. This suggests that the corresponding particles attract each other and collide. After the collision the particles diverge. When $t$ increases the \emph{world lines} become closer to the straight lines corresponding to the asymptotic solutions~\eqref{eq:asg} and~\eqref{eq:akdv} when $t \to \infty$. So the interaction between particles results in exchange of energy and phase shift which can be calculated as the distance between corresponding asymptotes.\\
Case 2. presents the interaction of the particles of the same type. When the values of $|t|$ are large and $t < 0$ the lines are close to the straight lines corresponding to the asymptotic solutions~\eqref{eq:asg} and~\eqref{eq:akdv} when $t \to -\infty$. Then after some convergence the lines diverge and do not intersect. This suggests that the corresponding particles repulse each other. When $t$ increases the \emph{world lines} become closer to the straight lines corresponding to the asymptotic solutions~\eqref{eq:asg} and~\eqref{eq:akdv} when $t \to \infty$. So as in the case 1. the interaction between particles results in exchange of energy and phase shift but without collision.\\
Case 3. corresponds to the periodical solutions that can be interpreted as \emph{bound state} of two particles of different types. This is similar to the "breathers" in case of classical soliton solutions of NIDE. Dynamics of the \emph{bound state} is similar to the dynamics of the "free" particle: particles oscillate around common center that propagates with the speed $v$ that can be calculated as $v = 1 / (\Re \omega_1)^2$.\\
When $N > 2$ it's not possible to calculate \emph{singularity lines} explicitly so numerical methods (i.e. finding the zeros of the transcendental functions $\delta_k(x, t), \; k = 1, 2$) should be applied. Nevertheless, there are some very interesting global properties of the \emph{singularity lines} that can be derived and investigated in details. \\
It's worth noting that in quantum mechanics the problem of studying a gas of one-dimensional Bose particles interacting via delta-function potential reduces to investigation of the $Schr\ddot{o}dinger$ equation
\begin{equation}\label{eq:bose}
\left( -\sum_{i = 1}^N {\left( \partial ^2 / \partial x_i^2 \right)} + 2 c \sum_{i, j = 1}^N {\delta(x_i - x_j)}\right) \psi = E \psi
\end{equation}
with boundary conditions
\begin{equation}\label{eq:bose1}
\left. \left( \frac{\partial}{\partial x_j} - \frac{\partial}{\partial x_k} \right) \psi \right|_{x_i = x_{k+}} - \left. \left( \frac{\partial}{\partial x_j} - \frac{\partial}{\partial x_k} \right) \psi \right|_{x_i = x_{k-}} = \left. 2 c \psi \right|_{x_i = x_k},
\end{equation}
i.e. $\psi$ is continuous whenever two particles touch, but the jump in the derivative of $\psi$ is $2 c$ (see for example \cite{qm1, qm2}). In this context our case can be considered as a generalization of the problem~\eqref{eq:bose},~\eqref{eq:bose1} and reduces to the one when $\omega_i >> \omega_j, \; i > j, \; 1 \leq i, j \leq N$. In this case in the limit $(\omega_i - \omega_j) \to \infty, \; i > j, \; 1 \leq i, j \leq N$ the region of particles' interaction collapses to the point.\\     
From~\eqref{eq:slines} it follows that \emph{singularity lines} satisfy the system of equations
\begin{equation} \label{eq:slinesde}
\begin{cases}
    \frac{\mathrm d x_i(t)}{\mathrm d t} = -\left ( \frac{\partial \delta_1(x, t)}{\partial t} / \frac{\partial \delta_1(x, t)}{\partial x} \right )_{x = x_i(t)}, \; i = 1, 2, \ldots, l;\\
    \frac{\mathrm d x_i(t)}{\mathrm d t} = -\left ( \frac{\partial \delta_2(x, t)}{\partial t} / \frac{\partial \delta_2(x, t)}{\partial x} \right )_{x = x_i(t)}, \; i = l + 1, l + 2, \ldots, N.\\
\end{cases}
\end{equation}
Let's introduce quantities
\begin{equation} \label{eq:slinesaa}
\begin{cases}
    p_i = \frac{\Re \Theta(\omega_i)}{\Re \omega_i};\\
    q_i = -p_i t + \ln|C_i| / \Re \omega_i, \; 1 \leq i \leq N,\\
\end{cases}
\end{equation}
where $\Theta(x)$ is defined in~\eqref{eq:sys310}.
In \cite{dis} the following theorem is proved
\begin{thm}
System~\eqref{eq:slinesde} is completely integrable Hamiltonian system with the Hamiltonian
\begin{equation} \label{eq:Ham}
H = \frac{1}{2}\sum_{i = 1}^{N}{p_i^2},
\end{equation}
and quantities~\eqref{eq:slinesaa} are action-angle variables for this system.
\end{thm}
\emph{Proof.} We just need to verify the validity of the identity
\begin{equation}
\frac{\mathrm d x}{\mathrm d t} = \left \{ x, H \right \},
\end{equation} 
where $\{x, H\}$ is the Poisson bracket defined by
$$
\{f, g\} = \sum_{i = 1}^{N} \left( \frac{\partial f}{\partial q_i} \frac{\partial g}{\partial p_i} - \frac{\partial f}{\partial p_i} \frac{\partial g}{\partial q_i}\right).
$$
Indeed, from~\eqref{eq:slinesaa} and~\eqref{eq:Ham} it follows that 
\begin{equation} \label{eq:dH}
\frac{\partial H}{\partial q_i} = 0, \; \frac{\partial H}{\partial p_i} = p_i; 1 \leq i \leq N.
\end{equation}
On the other hand, we have
\begin{equation} \label{eq:dxdq}
\frac{\partial x}{\partial q_i} = \frac{\partial \delta(x, t)}{\partial q_i} / \frac{\partial \delta(x, t)}{\partial x},
\end{equation}
where
\[ \delta(x, t) = 
\begin{cases}
     \delta_1(x, t), & \; \text{if } 1 \leq i \leq l;\\
    \delta_2(x, t), & \; \text{if } l < i \leq N.\\
\end{cases}
\]\\
Substituting 
$$
\frac{\partial \delta(x, t)}{\partial q_i} = \frac{\partial \delta(x, t)}{\partial t} / \frac{\partial q_i}{\partial t} = -\frac{\partial \delta(x, t)}{\partial t} / p_i
$$
into~\eqref{eq:dxdq} and combining with~\eqref{eq:dH} we obtain~\eqref{eq:slinesde}. $\square$

\begin{Cy} The total energy of the system of particles with the dynamics described by~\eqref{eq:slinesde} is an integral of motion.
\end{Cy}
\begin{Cy} System~\eqref{eq:slinesde} carries a complete information about the solution $\xi(x, t)$ of NIDE (it is contained in the sets $A = \left \{\alpha_{k,0} \right \}_{k = i}^N$ and $\Omega = \left \{\omega \right \}_{k = i}^N$). Having this information and using~\eqref{eq:Gamma1}-~\eqref{eq:Delta2} one can reconstruct the solutions.
\end{Cy}
\begin{Rk}
Equations~\eqref{eq:slines} solve N-body problem with a special potential.
\end{Rk}
\begin{Rk}Considered NIDE themselves can be formulated in terms of Hamiltonian systems in infinite dimensional space so we face a hierarchy of the Hamiltonian systems: infinite dimensional system generates the finite dimensional one.
\end{Rk}
Even though \textbf{Theorem 3.11.} states an important and powerful result, it's not constructive in a sense that it describes dynamics of the system implicitly: on the right hand side of the equations~\eqref{eq:slinesde} one cannot distinguish one \emph{singularity line} from another. It would be interesting to get some more detailed information about the behavior of \emph{singularity lines}. Using the results of \textbf{Section 2} we obtain the parametrization of the \emph{singularity lines} and derive differential equations for the parameters. In order to do this we need a simple result obtained in \cite{heinig3}: connection between the determinants of paired Cauchy and paired Vandermonde matrices. 
\begin{Dn}
Matrix $S$ is called paired Cauchy (PC) matrix if it (or its transposed) has the following block representation
$$
S = 
 \begin{bmatrix}
  S_1 \\
  S_2  \\ 
 \end{bmatrix}
$$
where $S_k, \; k = 1, 2$ are pure Cauchy matrices.
\end{Dn}
For example, matrix $S(x, t)$ represented by formula~\eqref{eq:sys321} is PC matrix.
\begin{Dn}
Matrix $V$ is called paired Vandermonde (PV) matrix if it (or its transposed) has the following block representation
$$
V = 
 \begin{bmatrix}
  V_1 \\
  V_2  \\ 
 \end{bmatrix}
$$
where $V_k, \; k = 1, 2$ are pure Vandermonde matrices.
\end{Dn}
For example, matrices $V_k(x, t),  k = 1, 2$ whose determinants $\Delta_k(x, t)$ are represented by formulas~\eqref{eq:Gamma2},~\eqref{eq:Gamma3} are PV matrices. The following assertion is true
\begin{La} Let the sets of numbers $\left \{\omega_i \right \}_{i = 1}^{m + n}$ and $\left \{\alpha_i \right \}_{i = 1}^{m + n}$ be such that $\omega_i \ne \omega_k, \; \alpha_i \ne \alpha_k; \; i \ne k$ and $\omega_i \ne -\alpha_k, 1 \leq i, k \leq m + n$. Define PC matrix $S$ by
$$
    S = \begin{bmatrix}
\frac{1}{\omega_1 + \alpha_1} & \frac{1}{\omega_2 + \alpha_1} & \cdots & \frac{1}{\omega_{m + n} + \alpha_1} \\
\cdots & \cdots & \cdots & \cdots\\
\frac{1}{\omega_1 + \alpha_m} & \frac{1}{\omega_2 + \alpha_m} & \cdots & \frac{1}{\omega_{m + n} + \alpha_m} \\
& & & \\
\frac{\gamma_1}{\omega_1 + \alpha_{m + 1}} & \frac{\gamma_2}{\omega_2 + \alpha_{m + 1}} & \cdots & \frac{\gamma_{m + n}}{\omega_{m + n} + \alpha_{m + 1}} \\
\cdots & \cdots & \cdots & \cdots\\
\frac{\gamma_1}{\omega_1 + \alpha_{m + n}} & \frac{\gamma_2}{\omega_2 + \alpha_{m + n}} & \cdots & \frac{\gamma_{m + n}}{\omega_{m + n} + \alpha_{m + n}} \\
\end{bmatrix}
$$
and VC matrix $V$ by
$$
    V = \begin{bmatrix}
1 & 1 & \cdots & 1 \\
\omega_1 & \omega_2 & \cdots & \omega_{m + n} \\
\cdots & \cdots & \cdots & \cdots\\
\omega_1^{m - 1} & \omega_2^{m - 1} & \cdots & \omega_{m + n}^{m - 1} \\
& & & \\
\epsilon_1 & \epsilon_2 & \cdots & \epsilon_{m + n} \\
\omega_1 \epsilon_1 & \omega_2 \epsilon_2 & \cdots & \omega_{m + n} \epsilon_{m + n} \\
\cdots & \cdots & \cdots & \cdots\\
\omega_1^{n - 1} \epsilon_1 & \omega_2^{n - 1} \epsilon_2& \cdots & \omega_{m + n}^{n - 1} \epsilon_{m + n} \\
\end{bmatrix},
$$
then
\begin{equation}
\det S = \frac{\prod_{1 \leq l < k \leq m}{(\alpha_k - \alpha_l)} \prod_{m + 1 \leq j < i \leq m + n}{(\alpha_i - \alpha_j)}}{\prod_{1 \leq k \leq m + n; 1 \leq i \leq m}{(\omega_k + \alpha_i)}} \det V,
\end{equation}
where
$$
\epsilon_k = \gamma_k \prod_{1 \leq i \leq m; m + 1 \leq j \leq m + n }{(\omega_k + \alpha_i) / (\omega_k + \alpha_j)}; \; 1 \leq k \leq m + n.
$$
\end{La}
The proof is based on the application of the Laplace rule to the calculation of the determinants and the properties of the determinants of pure Cauchy and Vandermonde matrices. It's a straightforward but bulky calculation and will be skipped (we refer the interested reader to \cite{heinig3} for the full proof; also in \cite{heinig3} one can find more links between different types of \emph{structured matrices}).\\
Combining results obtained in \textbf{Section 2} (\textbf{Theorem 2.1.}), formulas~\eqref{eq:Gamma}-~\eqref{eq:sys325} and \textbf{Lemma 3.32}, it's easy to see that the following statement is valid.
\begin{thm}
\begin{equation} 
\det S(x, t) = 0 \iff
\begin{cases}
    \delta_1(x, t) = 0,\\
    \delta_2(x, t) = 0.\\
\end{cases}
\end{equation}
\end{thm}
Taking into account \textbf{Remark 2.2.} we see that \emph{singularity lines} of the solutions $\xi(x, t)$ of NIDE are parametrized by the coefficients $\omega_k, \; 1 \leq k \leq N$ of some polynomials. As an example consider real solutions $\xi(x, t)$ for the case $N = 2$ (this example was also considered in \cite{heinig3}). The parametrizing polynomials $f_k(z), \: k = 1, 2$ in this case are of order one: $ f_1(z) = z - p, \: f_2(z) = z + p; \; p = \bar{p}$.

\begin{Ee}
Let $\omega_k = \bar{\omega}_k, \; \alpha_k = \bar{\alpha}_k, \; k = 1, 2$ and $\omega_2 > \omega_1$. Consider two cases:
\begin{enumerate}
\item $C_k > 0, k = 1, 2$,
\item $C_1 > 0, \; C_2 < 0$, 
\end{enumerate}
where
$$
C_k =  \frac {(\omega_k - \alpha_{1, 0})(\omega_k - \alpha_{2, 0})} {(\omega_k + \alpha_{1, 0})(\omega_k + \alpha_{2, 0})}, \; k = 1, 2.
$$
In the first case two \emph{singularity lines} $x_k(t), \: k = 1, 2$ solve the systems
\begin{equation} \label{eq:param1}
\begin{cases}
    \exp{2(\omega_k x + \Theta(\omega_k) t)} = C_k (\omega_k + p) / (\omega_k - p), \\
    -\exp{2(\omega_k x + \Theta(\omega_k) t)} = C_k (\omega_k + p) / (\omega_k - p); \; k = 1, 2. \\
\end{cases}
\end{equation}
One line ($L_1$) corresponds to the values of $p$ in the interval ${]{-\omega_1}, \omega_1[}$ and for the other one ($L_2$), $p \in {]{-\infty}, {-\omega_2}[} \cup {]\omega_2, \infty[}$.
Here $\Theta(x)$ is defined in~\eqref{eq:sys310}.
Solving~\eqref{eq:param1} with respect to $x$ and $t$ one gets
\begin{equation} \label{eq:param12}
    x = \frac{\vartheta_1(p) \Theta(\omega_2) - \vartheta_2(p) \Theta(\omega_1)}{\omega_1 \Theta(\omega_2) - \omega_2 \Theta(\omega_1)}, \: t = \frac{\vartheta_2(p) \omega_1 - \vartheta_1(p) \omega_2}{\omega_1 \Theta(\omega_2) - \omega_2 \Theta(\omega_1)},
\end{equation}
where
\begin{equation}\label{eq:vtheta}
	\vartheta_k(d) = \frac{1}{2}\ln{\left( C_k (\omega_k + p) / (\omega_k - p) \right)}, \; k = 1, 2.
\end{equation}
This case corresponds to the "attracting" particles (interaction between different types of particles) considered in \textbf{Example 3.25.} Case 1. Lines $L_1$ and $L_2$ intersect each other. From~\eqref{eq:param1}-~\eqref{eq:vtheta} it follows that coordinates $(x_0, t_0)$ of the intersection point satisfy the relation
\begin{equation}
    x_0 = \frac{1}{2} \frac{\ln{C_1} \Theta(\omega_2) - \ln{C_2} \Theta(\omega_1)}{\omega_1 \Theta(\omega_2) - \omega_2 \Theta(\omega_1)}, \: t_0 = \frac{1}{2} \frac{\ln{C_2} \omega_1 - \ln{C_1} \omega_2}{\omega_1 \Theta(\omega_2) - \omega_2 \Theta(\omega_1)}.
\end{equation}
This is achieved by setting $p = 0$ in case of line $L_1$ and $p \to \pm \infty$ in case of line $L_2$. By calculating derivative
$$
\frac{\mathrm d x(t)}{\mathrm d t} = \frac{\omega_1 \Theta(\omega_2)(\omega_2^2 - p^2) - \omega_2 \Theta(\omega_1)(\omega_1^2 - p^2)}{\omega_1 \omega_2 (\omega_1^2 - \omega_2^2)}
$$
for both lines, and setting $p = 0$ in case of $L_1$ and $p \to \pm \infty$ in case of $L_2$ we see that 
\begin{equation}
\left.\frac{\mathrm d x(t)}{\mathrm d t}\right|_{p = 0} = \begin{cases}
    0 & \text{in case of SHG equation} \; (\Theta(x) = 1/x), \\
    -(\omega_1^2 + \omega_2^2) & \text{in case of mKdV equation} \; (\Theta(x) = -x^3);\\
\end{cases}
\end{equation}
and 
$$
\left.\frac{\mathrm d x(t)}{\mathrm d t}\right|_{p \to \pm \infty} \to -\infty.
$$
So in case of SHG equation \emph{singularity lines} intersect at the angle $\pi / 2$.\\
In the second case \emph{singularity lines} solve the systems
\begin{equation} \label{eq:param2}
\begin{cases}
    -\exp{2(\omega_1 x + \Theta(\omega_1) t)} = C_1 (\omega_1 + p) / (\omega_1 - p); \\
    \exp{2(\omega_2 x + \Theta(\omega_2) t)} = |C_2| (\omega_2 + p) / (\omega_2 - p). \\
\end{cases}
\end{equation}
Corresponding intervals for the parameter $p$ are $]{-\omega_2}, {-\omega_1}[$ and $]\omega_1, \omega_2[$. Solutions of the systems~\eqref{eq:param2} have the same representation~\eqref{eq:param12} as in case 1. but this time they correspond to the "repulsing" particles (interaction between same type particles) considered in \textbf{Example 3.25.} Case 2.
\end{Ee}
Consider now a general case of parametrization of the \emph{singularity lines} of the real solutions $\phi(x, t)$ of SHG equation (MKdV and NSE equations can be investigated in the similar manner). Parametrizing polynomials $Q_{1, 2}(x)$ from~\eqref{eq:param26} in this case are of order $N - 1$. Taking into account symmetries imposed on the sets $\left \{\omega_i \right \}_{i = 1}^{N}$, $\left \{\alpha_{i, 0} \right \}_{i = 1}^{N}$, polynomials $Q_{1, 2}(x)$ can be represented as
$$
Q_{1, 2}(x) = \prod_{i = 1}^{N - 1}{(x \pm p_i)},
$$ 
where $p_i = \bar{p}_i, \; 1 \leq i \leq N - 1$ so the parametrization is performed by the roots $\left \{\pm p_i \right \}_{i = 1}^{N - 1}$ of the polynomials $Q_{1, 2}(x)$ where $Q_2(x) = (-1)^{N - 1}Q_1(-x)$. The parametrization takes the form
\begin{equation} \label{eq:parametrization}
\sign(C_k) \exp\left [{2 (\omega_k x + \frac{1}{\omega_k} t )}\right ] = C_k \prod_{i = 1}^{N - 1}{\frac{\omega_k + p_i}{\omega_k - p_i}}, \; 1 \leq k \leq N.
\end{equation}
Now we prove the following theorem
\begin{thm}
 Let the sets of numbers $\left \{\omega_i \right \}_{i = 1}^{N}$, $\left \{\alpha_{i, 0} \right \}_{i = 1}^{N}$ be such that $\omega_i = \bar{\omega}_i, \: \alpha_i = \bar{\alpha}_i; \: 1 \leq i \leq N; \: \omega_i \ne \omega_k, \; \alpha_i \ne \alpha_k; \; i \ne k$ and $\omega_i \ne \alpha_k, 1 \leq i, k \leq N$, then parameters $\left \{p_i \right \}_{i = 1}^{N - 1}$ considered as functions of $t$, satisfy the nonlinear system of differential equations
\begin{equation} \label{eq:paramde}
\frac{\mathrm d p_k(t)}{\mathrm d t} = (-1)^N \frac{\prod_{1 \leq i \leq N -1,  i \neq k}{p_i^2(t)} \prod_{1 \leq i \leq N}{(\omega_i^2 - p_k^2(t))}}{\prod_{1 \leq i \leq N}{\omega_i^2} \prod_{1 \leq i \leq N - 1,  i \neq k}{(p_k^2(t) - p_i^2(t))}}, \: 1 \leq k \leq N - 1.
\end{equation}
\end{thm}
\emph{Proof.} Suppose for definiteness that in~\eqref{eq:parametrization} $C_k > 0, \: 1 \leq k \leq N$. Then fix some index $k$ (without loss of generality we can take $k = 1$) and calculate $x$
\begin{equation} \label{eq:calcx}
x = \frac{1}{\omega_1}\left( \frac{1}{2} \sum_{i = 1}^{N - 1}{\ln{\frac{\omega_1 + p_i}{\omega_1 - p_i}} - \frac{1}{\omega_1} t + \frac{1}{2} \ln C_1}\right).
\end{equation}
To simplify the notations, the dependence of $p_k(t)$ from $t$ is omitted. 
Substituting~\eqref{eq:calcx} into~\eqref{eq:parametrization} for all $k > 1$ we come up with the system
\begin{equation} \label{eq:calcx2}
\begin{split}
F_k(x, t, p) & \equiv \frac{\omega_k}{\omega_1} \left ( \frac{1}{2} \sum_{i = 1}^{N - 1}{\ln{\frac{\omega_1 + p_i}{\omega_1 - p_i}}}  + \frac{1}{\omega_1} t - \frac{1}{2} \ln C_1\right ) \\
             & + \frac{t}{\omega_k} + \frac{1}{2} \ln C_k - \frac{1}{2}\sum_{i = 1}^{N - 1}{\ln{\frac{\omega_k + p_i}{\omega_k - p_i}}} = 0,\\
\end{split}
\end{equation}
where $2 \leq k \leq N$.
After differentiating~\eqref{eq:calcx2} with respect to $t$ to obtain
$$
\frac{\partial F_k(x, t, p)}{\partial t} + \sum_{i = 1}^{N - 1} \left( \frac{\partial F_k(x, t, p)}{\partial p_i}\frac{\mathrm d p_i}{\mathrm d t} \right) = 0
$$
and substituting the derivatives    
$$
\frac{\partial F_k(x, t, p)}{\partial p_i} =  \frac{\omega_k \left(\omega_k^2 - \omega_1^2 \right)}{(\omega_1^2 - p_i^2)(\omega_k^2 - p_i^2)}
$$ 
$$
\frac{\partial F_k(x, t, p)}{\partial t} = -\frac{\omega_k^2 - \omega_1^2}{\omega_1^2 \omega_k}
$$    
we come up with the system of linear equations with respect to the derivatives $ \mathrm d p_i / \mathrm d t$
\begin{equation} \label{eq:system}
\sum_{i = 1}^{N - 1}{\frac{1}{(\omega_1^2 - p_i^2)(\omega_k^2 - p_i^2)} \frac{\mathrm d p_i}{\mathrm d t}} = \frac{1}{\omega_k^2 \omega_1^2}, \: 2\leq k \leq N.
\end{equation}
The determinant
$$
\Delta = \det \left \{ \frac{1}{(\omega_1^2 - p_i^2)(\omega_k^2 - p_i^2)} \right \}_{2 < k \leq N, \: 1 \leq i \leq N - 1}
$$
of the matrix coefficient of the system~\eqref{eq:system} can be expressed as
$$
\Delta = \prod_{i = 1}^{N - 1}{(\omega_1^2 - p_i^2)^{-1}}\det \left \{ \frac{1}{\omega_k^2 - p_i^2} \right \}_{2 < k \leq N, \: 1 \leq i \leq N - 1}, 
$$
where the second factor on the right hand side is the determinant of the Cauchy matrix. So the matrix coefficient is non-singular and the system~\eqref{eq:system} has a unique solution. Using Cramer's rule, after simple manipulations with explicit formulas for the Cauchy matrix determinants, we arrive at~\eqref{eq:paramde}. $\square$ \\\\
Analysis of the system~\eqref{eq:paramde} is quite non-trivial and will be carried out in subsequent publications. It's easy to see, though, that the system~\eqref{eq:paramde} has some important properties that will be useful in our further considerations. For example, equations do not depend on the set $\left \{\alpha_{i, 0} \right \}_{i = 1}^{N}$ which makes it easier to address inverse problems. Also, the system~\eqref{eq:paramde}, as opposed to ~\eqref{eq:slinesde}, allows to distinguish between different \emph{singularity lines}. This is based on the following observation. Let's assume that the set $\left \{\omega_i \right \}_{i = 1}^{N}$ is such that $\omega_i = \bar{\omega}_i, \: 1 \leq i \leq N$. Because of the symmetry, without loss of generality, we can assume that $\omega_i > 0, \: 1 \leq i \leq N$ and numbers $\omega_i$ are enumerated such that $\omega_i > \omega_k, \: i > k$. In this setup real axes ${]-\infty, \infty[}$ is divided into non-overlapping intervals $\Omega_{2N} = {]-\infty, -\omega_N[} \;  \cup \; {]\omega_N, \infty[}, \; \Omega_{2N - 1} = {]-\omega_N, -\omega_{N - 1}[}, \; \ldots, \; \Omega_{N} = {]-\omega_1, \omega_1[}, \; \ldots, \; \Omega_{1} = {]\omega_{N - 1}, \omega_N[}$. There is one-to-one correspondence between initial values of the parameters $p_i(t_0), \: 1 \leq i \leq N - 1$ and intervals $\Omega_k, \: 1 \leq k \leq 2N$ such that the values $p_i(t_0)$ can only belong to the different intervals and over time the initial mapping doesn't change (the proof of this statement in general setup requires a non-trivial analysis of the system~\eqref{eq:paramde} and will be addressed in the subsequent publication). So the particular \emph{singularity line} $L_k$ is characterized by the particular function $p_k(t)$ taking values from the particular interval $\Omega_k; \: 1 \leq k \leq N$.
We illustrate the above statements by a simple example.
\begin{Ee} Consider SHG equation and let $N = 2$ and $\omega_i = \bar{\omega}_i, \: \omega_i > 0, \: \alpha_{i, 0} = \bar{\alpha}_{i, 0}; \: i = 1, 2; \: \omega_2 > \omega_1$. In this case we have one parameter $p(t)$ and system~\eqref{eq:paramde} takes the form
\begin{equation}\label{eq:paramde2}
\frac{\mathrm d p(t)}{\mathrm d t} = \frac{p^2(t) (\omega_1^2 - p^2(t))(\omega_2^2 - p^2(t))}{\omega_1^2 \omega_2^2}.
\end{equation} 
We  also supply a special initial condition $p^*(t_0^*) = p_0^*$. Equation~\eqref{eq:paramde2} can be easily integrated giving a general solution
\begin{equation}\label{eq:paramsol2} 
t - t_0^* = \frac{\omega_1^2 \omega_2^2}{2 (\omega_2^2 - \omega_1^2)}\left({\frac{1}{\omega_1} \ln\left|{\frac{\omega_1 + p(t)}{\omega_1 - p(t)}}\right| - \frac{1}{\omega_2} \ln\left|{\frac{\omega_2 + p(t)}{\omega_2 - p(t)}}\right|}\right).
\end{equation}
It follows from~\eqref{eq:paramsol2} that $p_0^*$ should satisfy the consistency condition
\begin{equation}\label{eq:consist}
\left|\frac{\omega_1 + p_0^*}{\omega_1 - p_0^*}\right|^{\omega_2} = \left|\frac{\omega_2 + p_0^*}{\omega_2 - p_0^*}\right|^{\omega_1}.
\end{equation} 
Equation~\eqref{eq:consist} has four distinct solutions:
\begin{enumerate}
	\item $p_{0, 1}^* = 0;$
	\item $p_{0, 2}^* = \pm \infty;$
	\item $p_{0, 3}^* = p^*;$
	\item $p_{0, 4}^* = -p^*;$
\end{enumerate}
where $p^* > 0, \: p^* \in ]\omega_1, \omega_2[$. So each value of $p_0^*$ belongs to one of the intervals $]-\infty, -\omega_2[ \; \cup \; ]\omega_2, \infty[, \; ]-\omega_2, -\omega_1[, \; ]-\omega_1, \omega_1[, \; ]\omega_1, \omega_2[$.\\
In case $p_0^* = 0$ we have $p(t) \in ]-\omega_1, \omega_1[$ and corresponding \emph{singularity line} $x(t)$ solves each of the equations
\begin{equation}\label{eq:sleq1} 
\exp{2(\omega_k x(t) + t / \omega_k)} = C_k (\omega_k + p(t)) / (\omega_k - p(t)), \: k = 1, 2.
\end{equation}
It is required in this case that $C_k > 0, \: k = 1, 2$.\\
In case $p_0^* = \pm \infty$ we have $p(t) \in ]-\infty, -\omega_2[ \; \cup \; ]\omega_2, \infty[$ and corresponding \emph{singularity line} $x(t)$ solves each of the equations
\begin{equation}\label{eq:sleq2}
-\exp{2(\omega_k x(t) + t / \omega_k)} = C_k (\omega_k + p(t)) / (\omega_k - p(t)), \: k = 1, 2.
\end{equation}
It is also required in this case that $C_k > 0, \: k = 1, 2$. Considered cases (1. 2.) correspond to the case of "attracting" particles discussed in \textbf{Example 3.25.} Case 1. \\
Analogously, consider the cases 3. and 4. In case $p_0^* = p^*, \; p^* > 0, \: p^* \in ]\omega_1, \omega_2[$ we have $p(t) \in ]\omega_1, \omega_2[$ and corresponding \emph{singularity line} $x(t)$ solves each of the equations
\begin{equation}\label{eq:sleq3}
-\exp{2(\omega_1 x(t) + t / \omega_k)} = C_1 (\omega_1 + p(t)) / (\omega_1 - p(t)),
\end{equation}
\begin{equation}\label{eq:sleq4}
\exp{2(\omega_2 x(t) + t / \omega_2)} = |C_2| (\omega_2 + p(t)) / (\omega_2 - p(t)).
\end{equation}
It is required in this case that $C_1 > 0, \; C_2 < 0$.\\
In case $p_0^* = -p^*$ we have $p(t) \in ]-\omega_2, -\omega_1[$ and corresponding \emph{singularity line} $x(t)$ solves each of the previous equations. It is also required in this case that $C_1 > 0, \; C_2 < 0$. This corresponds to the case of "repulsing" particles discussed in \textbf{Example 3.25.} Case 2.
\end{Ee}
From the considered example it follows that the triplets $(x_{0, i}^*, t_0^*, p_{0, i}^*), \; 1 \leq i \leq 4$ are completely determined by the sets $\left\{\omega_1, \omega_2\right\}, \; \left\{\alpha_{1, 0}, \alpha_{2, 0}\right\}$. Taking into account~\eqref{eq:consist} it's easy to calculate $t_0^*$ and $x_{0, i}^*, \; 1 \leq i \leq 4$:
\begin{equation}\label{eq:sleqt0}
t_0^* = \frac{\omega_1 \omega_2 (\omega_2 \ln |C_1| - \omega_1 \ln |C_2| )}{2 (\omega_2^2 - \omega_1^2)},
\end{equation}
\begin{equation}\label{eq:sleqx0}
x_{0, i}^* = \frac{1}{2 \omega_1}\left( \ln{\left|\frac{\omega_1 + p_{0, i}}{\omega_1 - p_{0, i}}\right|} +\frac{\omega_1 (\omega_2 \ln |C_1| - \omega_1 \ln |C_2| )}{\omega_2^2 - \omega_1^2}\right), \; 1 \leq i \leq 4.
\end{equation}
Thus given the sets $\left\{\omega_1, \omega_2\right\}, \; \left\{\alpha_{1, 0}, \alpha_{2, 0}\right\}$ the alternative method of construction of the \emph{singularity lines} is reduced to the following steps:
\begin{enumerate}
	\item Step1: From~\eqref{eq:consist} calculate $p_{0, i}^*, \; \; 1 \leq i \leq 4$ and $t_0^*$ from~\eqref{eq:sleqt0};  
	\item Step2: Solve differential equation~\eqref{eq:paramde2} with initial data $(t_0^*, p_{0, i}^*)$ to obtain $p_i(t), \; 1 \leq i \leq 4$;
	\item Step3: Substitute $p_i(t)$ into corresponding equation~\eqref{eq:sleq1}-~\eqref{eq:sleq4} to obtain $x_i(t), \; 1 \leq i \leq 4$.
\end{enumerate}
Described methodology is also valid in general case but the solution of the system~\eqref{eq:paramde} cannot be constructed in closed form and should involve numerical methods.\\
As it was pointed out before, \emph{singularity lines} contain full information about the PE(N) solutions of NIDE. In this respect it would be interesting to consider the following problem:

\begin{Pb} Given some information about singularity lines, restore the corresponding PE(N) solutions of NIDE.
\end{Pb}
We restrict ourselves to considering a special case of the \textbf{Problem 3.35} for the SHG equation when $N = 2$ (general case will be considered in further publications). In this case \textbf{Problem 3.35} is solved by the following assertion:
  
\begin{An}
The system (PE(N) solutions of NIDE) is characterized by the following data
\begin{equation} \label{eq:data}
\left\{t_0, \; \left. \frac{\mathrm d x_j^i(t)}{\mathrm d t^i}\right|_{t = t_0}\right\}, \; j = 1, 2; \: i = 0, 1, 2
\end{equation}
at some point $t_0 \in ]-\infty, \infty[$, and index $j$ enumerates singularity lines for the particular case ("attracting" (A-case) or "repulsing" (R-case)).
\end{An}
\emph{Proof.} To simplify the notations we adopt the following designations:
$$
\ddot{x}_j \equiv \left. \frac{\mathrm d x_j^2(t)}{\mathrm d t^2}\right|_{t = t_0}, \; \dot{x}_j \equiv \left. \frac{\mathrm d x_j(t)}{\mathrm d t}\right|_{t = t_0}, x_j \equiv x_j(t_0), \; p_j \equiv p_j(t_0); \; j = 1, 2.
$$
It suffice to show that given data~\eqref{eq:data}, one can uniquely recover the sets $\left\{\omega_1, \omega_2\right\}, \; \left\{\alpha_{1, 0}, \alpha_{2, 0}\right\}$. Indeed, differentiating equations~\eqref{eq:sleq1} -~\eqref{eq:sleq4} corresponding to the particular case, in the neighborhood of $t_0$ with respect to $t$ and using~\eqref{eq:paramde2} we obtain the following relations
\begin{equation} \label{eq:invsl1}
\omega_1^2 \omega_2^2 \frac{\mathrm d x_j(t)}{\mathrm d t} = -p_j^2 (t), \; j = 1, 2.
\end{equation}
Differentiating~\eqref{eq:invsl1} one more time with respect to $t$ and using again~\eqref{eq:paramde2}, results in 
\begin{equation} \label{eq:invsl2}
\omega_1^4 \omega_2^4 \frac{\mathrm d x_j^2(t)}{\mathrm d t^2} = -2 p_j(t) \left(p_j^2 (t) - \omega_1^2 \right) \left(p_j^2 (t) - \omega_2^2 \right), \; j = 1, 2.
\end{equation}
Setting $t = t_0$ in~\eqref{eq:invsl1} and~\eqref{eq:invsl2} we arrive at the system of four non-linear equations
\begin{equation} \label{eq:invsl3}
\begin{cases}
    \sigma_2(\omega^2) \dot{x}_j & = -p_j^2, \\
    \sigma_2^2(\omega^2) \ddot{x}_j & = -2 p_j (p_j^4 - \sigma_1(\omega^2) p_j^2 + \sigma_2(\omega^2)); \; j = 1, 2 \\
\end{cases}
\end{equation}
with respect to the unknowns $p_j$ and $\sigma_j(\omega^2), j = 1, 2$, where $\sigma_j(\omega^2)$ are symmetric functions of the set $\left\{\omega_1^2, \omega_2^2 \right\}: \sigma_1(\omega^2) = \omega_1^2 + \omega_2^2, \; \sigma_2(\omega^2) = \omega_1^2 \omega_2^2.$ Simple algebra gives the following quadratic equations for $p_j, \; j = 1, 2$: 
\begin{equation} \label{eq:invsl4}
\begin{cases}
    \dot{x}_1 (\dot{x}_1 - \dot{x}_2) p_2^2 - \frac{p_2}{2} \left( \frac{\dot{x}_1 \ddot{x}_2}{\dot{x}_2} + \sqrt{\frac{\dot{x}_2}{\dot{x}_1}} \ddot{x}_1 \right) + \dot{x}_1 - \dot{x}_2 = 0, \\
    \dot{x}_2 (\dot{x}_1 - \dot{x}_2) p_1^2 + \frac{p_1}{2} \left( \frac{\dot{x}_2 \ddot{x}_1}{\dot{x}_1} + \sqrt{\frac{\dot{x}_1}{\dot{x}_2}} \ddot{x}_2 \right) + \dot{x}_1 - \dot{x}_2 = 0. \\
\end{cases}
\end{equation}
Solving~\eqref{eq:invsl4} we obtain $p_j, \; j = 1, 2$. Then using the first of the equations~\eqref{eq:invsl3} we calculate $\sigma_2(\omega^2)$. Substituting $\sigma_2(\omega^2)$ into the second equation we find $\sigma_1(\omega^2)$. Calculating the roots of the polynomial $f(y) = y^2 - y \sigma_1(\omega^2) + \sigma_2(\omega^2)$ we find the values for $\omega_j, \; j = 1, 2$. Let's note that equations~\eqref{eq:invsl4} have two extra solutions that should be dropped by matching the values of $p_j, \; j = 1, 2$ and the intervals they fall into according to the considered case (A or R). Next we calculate $\left\{\alpha_{1, 0}, \alpha_{2, 0}\right\}$. It follows from~\eqref{eq:sleq1} -~\eqref{eq:sleq4} that symmetric functions $\sigma_j(\alpha_0), \; j = 1, 2$ satisfy the system of equations  
\begin{equation} \label{eq:invsl5}
\sigma_1(\alpha_0) \omega_j (1 + \kappa_j) - \sigma_2(\alpha_0) (1 - \kappa_j) = \omega_j^2 (1 - \kappa_j) , \; j = 1, 2,
\end{equation}
where
$$
\kappa_j = \sign(C_j) \exp{2 (\omega_j x_j + \frac{t_0}{\omega_j} -\frac{1}{2} \ln{\frac{\omega_j + p_j}{\omega_j - p_j}} )}, \; j = 1, 2.
$$
System~\eqref{eq:sleq1} has a unique solution from which we recover $\left\{\alpha_{1, 0}, \alpha_{2, 0}\right\}$ by solving quadratic equation $y^2 - y \sigma_1(\alpha_0) + \sigma_2(\alpha_0) = 0$.\\
In R-case when $t_0 = t_0^*$  we have the following symmetry relations
$$
p_1 = -p_2, \; \dot{x}_1 = \dot{x}_2, \; \ddot{x}_1 = -\ddot{x}_2,
$$
and from~\eqref{eq:invsl3} -~\eqref{eq:invsl4} it follows that
\begin{equation} \label{eq:invsl6}
p_j = \frac{\dot{x}_j}{\ddot{x}_j}, \; \sigma_2(\omega^2) = -\frac{a_j^2}{\dot{x}_j}, \; \sigma_1(\omega^2) = p_j^2 + \frac{p_j \ddot{x}_j}{2 \dot{x}_j^2} - \frac{1}{\dot{x}_j}.
\end{equation}
In~\eqref{eq:invsl6} index $j$ can be either 1 or 2.\\
In A-case when $t_0 = t_0^*$ the values of $p_j$ and the derivatives $\dot{x}_j, \ddot{x}_j, \; j = 1,2$ are trivial and don't carry any information so in this case the problem cannot be solved uniquely.
$\square$\\\\
We illustrate the methodology developed in \textbf{Assertion 3.36} by the numerical examples.
\begin{Ee}
Given the following data for the R-case:
$$
t_0 = -0.479042987; x_1 = 0.610504874; x_2 = -0.709437736; \dot{x}_1 = -0.713296278;
$$
$$
 \dot{x}_2 = -0.78498714; \ddot{x}_1 = 0.448732074; \ddot{x}_2 = -0.407660883.
$$
Calculation steps:
\begin{enumerate}
\item Step 1: Calculate $p_j, \; j = 1, 2$ using~\eqref{eq:invsl4}: $p_{1, 1} = -0.75418; \: p_{1, 2} = 1.68914; \: p_{2, 1} = 0.79117; \: p_{2, 2} = -1.77199;$
\item Step 2: Calculate $\sigma_j(\omega^2), \; j = 1, 2$: $\sigma_{1, 1}(\omega^2) = 1.6381478; \: \sigma_{1, 2}(\omega^2) = 5.0; \: \sigma_{2, 1}(\omega^2)= 0.7973978; \: \sigma_{2, 2}(\omega^2) = 4.0;$
\item Step 3: Calculate $\omega_j, \; j = 1, 2$: $\omega_{1, 1} - \text{complex}; \: \omega_{1, 2} = \pm 1.0; \: \omega_{2, 1} - \text{complex}; \: \omega_{2, 2} = \pm 2.0;$
\item Step 4: Verify the results: Values $\sigma_{1, 1}(\omega^2) = 1.6381478; \: \sigma_{2, 1}(\omega^2)= 0.7973978;$ and corresponding complex $\omega_{1, 1}$ and $\omega_{2, 1}$ should be dropped; $p_{2, 1} \in ]-2.0, -1.0[, \: p_{1, 2} \in ]1.0, 2.0[$;
\item Step 5: Calculate $\alpha_{0, j}, \; j = 1, 2$ using~\eqref{eq:invsl5}: $\alpha_{0, 1} = -0.71651, \: \alpha_{0, 2} = 1.116515$.
\end{enumerate}
\end{Ee}

Similarly, consider calculation steps for A-case.
\begin{Ee}
Given the following data for the A-case:
$$
t_0 = -0.550122329; x_1 = 0.012826762; x_2 = -0.201327063; \dot{x}_1 = -0.00003606;
$$
$$
 \dot{x}_2 = -6.285525817; \ddot{x}_1 = -0.006; \ddot{x}_2 = 319.9146357.
$$
Calculation steps:
\begin{enumerate}
\item Step 1: Calculate $p_j, \; j = 1, 2$ using~\eqref{eq:invsl4}: $p_{1, 1} = -13.24693417; \: p_{1, 2} = 0.01201; \: p_{2, 1} = -5.01419019; \: p_{2, 2} = 5530.611771;$
\item Step 2: Calculate $\sigma_j(\omega^2), \; j = 1, 2$: $\sigma_{1, 1}(\omega^2) = 30610058.78; \: \sigma_{1, 2}(\omega^2) = 5.0; \: \sigma_{2, 1}(\omega^2)= 4866365.591; \: \sigma_{2, 2}(\omega^2) = 4.0;$
\item Step 3: Calculate $\omega_j, \; j = 1, 2$: $\omega_{1, 1} - \text{complex}; \: \omega_{1, 2} = \pm 1.0; \: \omega_{2, 1} = \pm 5532.635804; \: \omega_{2, 2} = \pm 2.0;$
\item Step 4: Verify the results: Values $\sigma_{1, 1}(\omega^2) = 30610058.78; \: \sigma_{2, 1}(\omega^2)= 4866365.591;$ and corresponding $\omega_{1, 1}$ and $\omega_{2, 1}$ should be dropped; $p_{2, 1} \in ]-\infty, -2.0[, \: p_{1, 2} \in ]-1.0, 1.0[$;
\item Step 5: Calculate $\alpha_{0, j}, \; j = 1, 2$ using~\eqref{eq:invsl5}: $\alpha_{0, 1} = 0.0, \: \alpha_{0, 2} = 0.5$.
\end{enumerate}
\end{Ee}
In \textbf{Appendix} we present some of the examples of the behavior of the \emph{singularity lines} for the cases $N > 2$ obtained by numerical methods. These examples, on the one hand, reflect some general laws discussed previously e.g. asymptotic behavior when $|t| \to \infty$, the nature of the intersections of the \emph{singularity lines}; on the other hand, they introduce new effects admitting a non-trivial interpretation.\\
Figure 4 represents the interaction between three particles of the same type. As in the case of two particles of the same type, \emph{singularity lines} do not intersect (particles "repulse" each other). Also one can select regions where particles interact in pairs so complex interaction can locally be described in term of a simpler model ($N = 2$). This happens when corresponding values of the parameters $\omega$ are very distinguished from each other.\\
Figure 5 exhibits an interaction between three particles where two of them are of the same type and one is of different type. As in the previous example, one also can select regions where particles interact in pairs. \emph{Singularity lines} corresponding to the particles of different types, intersect (particles "attract" each other and "annihilate") and particles of the same type "repulse" each other.\\
Figure 6 demonstrates the interaction between "free" particle and a \emph{bound state}. "Free" particle "penetrates" into the \emph{bound state} and "knocks out" the one of the same type. A "knocked out" particle becomes "free" and the "knocking" particle gets "captured" by the particle of different type creating a new \emph{bound state}.\\
Figures 7 - 10 focus on the case $N = 4$. When parameters $\omega$ and $\alpha_0$ are
 real numbers the behavior of the \emph{singularity lines} is similar to the considered cases $N = 2, 3$ (Figures 7, 8). An interesting phenomena occurs in case of "bound states" interaction (Figures 9, 10). "Weak" interaction is presented on Figure 9. In this case "bound" states are interacting as "free" particles of the same type - they "repulse" each other. There is an exchange of energy between "bound states" but there is no exchange of individual particles. Figure 10 shows "strong" interaction between "bound states" with a complex exchange of particles between them. Closer look at the interaction region (inserts on the right and left hand sides) reveals a new type of interaction that couldn't be observed in cases of simpler systems ($N = 2, 3$): "generation" and "annihilation" of the virtual particles (encircled points of "generation" are marked by "G" and points of "annihilation" are marked by "A"). Some of the "virtual" particles exist for a short period of time and then "annihilate" with another "virtual" or "permanent" particle. But some of them "convert" to a "permanent" state replacing "annihilated" ones and form new "bound states" with "survived" particles of different types. One still can observe the exchange of energy between "bound states" on a large scale but tracking the behavior pattern of the individual particles in the presence of the "virtual" ones is quite problematic.   

\section{Appendix.}
Here we present the results of numerical calculations of the \emph{singularity lines} for the cases $N = 2, 3, 4$ and different combinations of the parameters $\omega$ and $\alpha_0$. \emph{Singularity lines} corresponding to the particles of the same type have the same color.
\begin{figure}[!ht]
  \centering
    \includegraphics[width=1.0\textwidth]{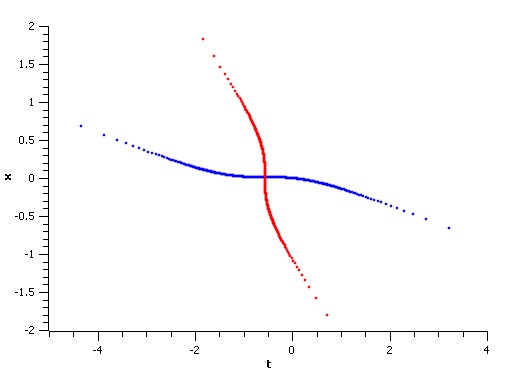}
		\label{fig:N2attract}
  \caption{Singularity lines, SHG equation, N = 2, "attracting" particles; parameters: $\omega_1 = 1.0, \; \omega_2 = 2.0, \; \alpha_1 = 0.0, \; \alpha_2 = 0.5.$}
\end{figure}
\begin{figure}[!ht]
  \centering
    \includegraphics[width=1.0\textwidth]{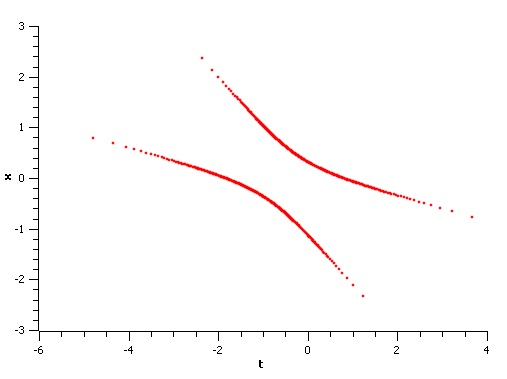}
		\label{fig:N2repulse}
  \caption{Singularity lines, SHG equation, N = 2, "repulsing" particles; parameters: $\omega_1 = 1.0, \; \omega_2 = 2.0, \; \alpha_1 = 1.116515, \; \alpha_2 = -0.71652$}
\end{figure}
\begin{figure}[!ht]
  \centering
    \includegraphics[width=1.0\textwidth]{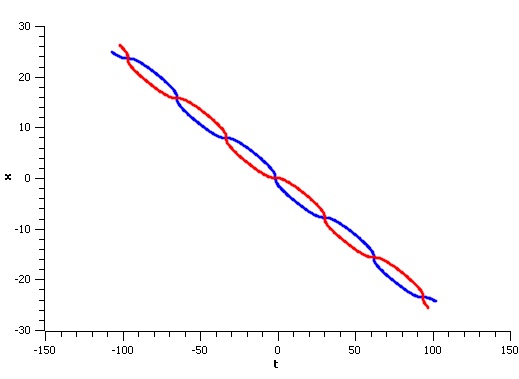}
		\label{fig:N2bs}
  \caption{Singularity lines, SHG equation, N = 2, "bound state"; parameters: $\omega_1 = 2.0 + \imath 0.2, \; \omega_2 = 2.0 - \imath 0.2, \; \alpha_1 = 0.1, \; \alpha_2 = 1.0$}
\end{figure}
\begin{figure}[!ht]
  \centering
    \includegraphics[width=1.0\textwidth]{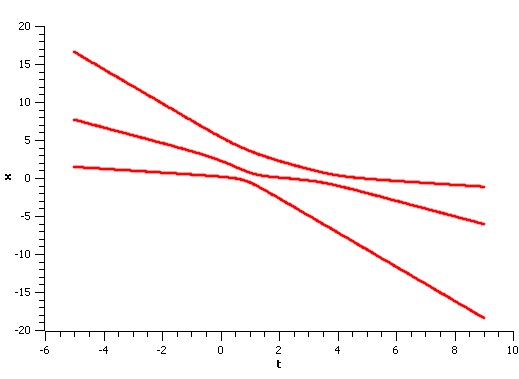}
		\label{fig:N3repulse}
  \caption{Singularity lines, SHG equation, N = 3, "repulsing" particles; parameters: $\omega_1 = 0.5, \; \omega_2 = 1.0, \; \omega_3 = 1.5, \; \alpha_1 = 0.2, \; \alpha_2 = 1.1, \; \alpha_3 = 1.3.$}
\end{figure}
\begin{figure}[!ht]
  \centering
    \includegraphics[width=1.0\textwidth]{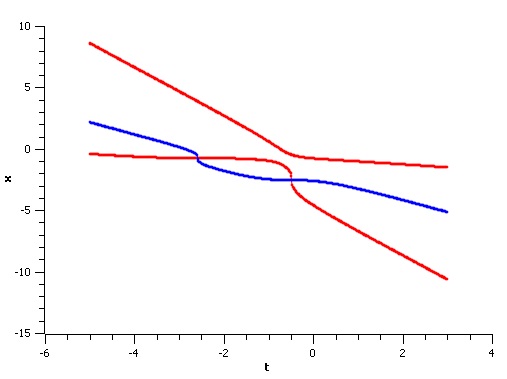}
		\label{fig:N3attract}
  \caption{Singularity lines, SHG equation, N = 3, "attracting" particles; parameters: $\omega_1 = 0.5, \; \omega_2 = 1.0, \; \omega_3 = 1.4, \; \alpha_1 = 0.4, \; \alpha_2 = 0.7, \; \alpha_3 = 1.2.$}
\end{figure}
\begin{figure}[!ht]
  \centering
    \includegraphics[width=1.0\textwidth]{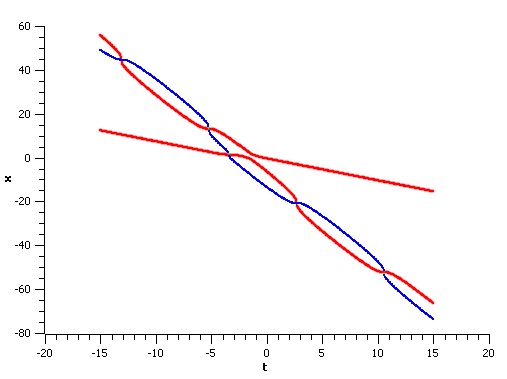}
		\label{fig:N3complex}
  \caption{Singularity lines, SHG equation, N = 3, "free" particle interacting with "bound state"; parameters: $\omega_1 = 1.0, \; \omega_2 = 2.0 + \imath 0.1, \; \omega_3 = 2.0 - \imath 0.1, \; \alpha_1 = 2.1, \; \alpha_2 = 2.2, \; \alpha_3 = 2.3.$}
\end{figure}
\begin{figure}[!ht]
  \centering
    \includegraphics[width=1.0\textwidth]{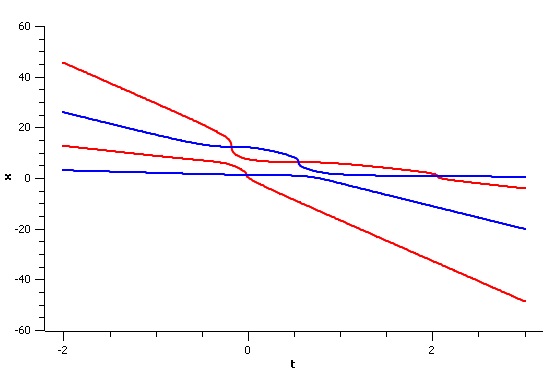}
		\label{fig:N4allpositive}
  \caption{Singularity lines, SHG equation, N = 4, two particles of the same type interacting with two particles of different type; parameters: $\omega_1 = 1.0, \; \omega_2 = 2.0, \; \omega_3 = 3.0, \; \omega_4 = 4.0, \; \alpha_1 = 0.2, \; \alpha_2 = 0.8, \; \alpha_3 = 2.2, \; \alpha_4 = 2.5.$}
\end{figure}
\begin{figure}[!ht]
  \centering
    \includegraphics[width=1.0\textwidth]{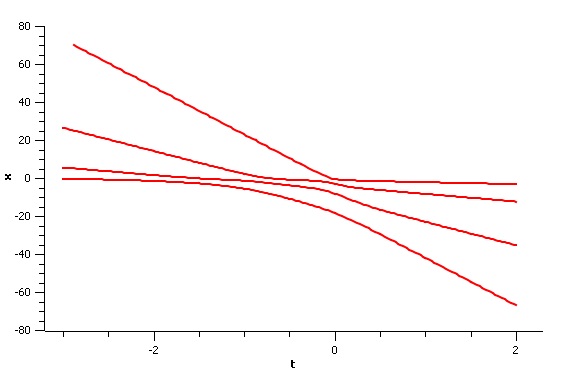}
		\label{fig:N4repulse}
  \caption{Singularity lines, SHG equation, N = 4, all particles of the same type; parameters: $\omega_1 = 1.0, \; \omega_2 = 2.0, \; \omega_3 = 3.0, \; \omega_4 = 4.0, \; \alpha_1 = 0.2, \; \alpha_2 = 1.3, \; \alpha_3 = 2.2, \; \alpha_4 = 3.5.$}
\end{figure}
\begin{figure}[!ht]
  \centering
    \includegraphics[width=1.0\textwidth]{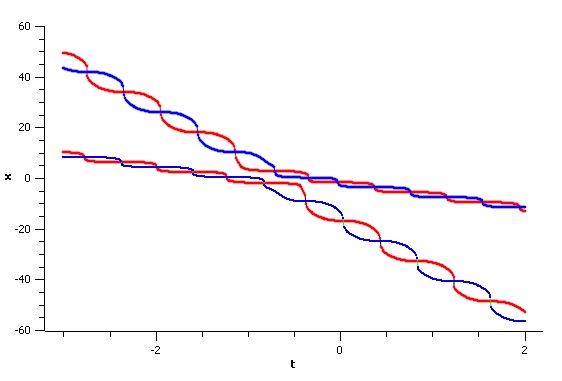}
		\label{fig:N4complex}
  \caption{Singularity lines, SHG equation, N = 4, "weak" interaction between two "bound states"; parameters: $\omega_1 = 1.0 + \imath 2.0, \; \omega_2 = 1.0 - \imath 2.0, \; \omega_3 = 4.0 + \imath 1.96, \; \omega_4 = 4.0 - \imath 1.96, \; \alpha_1 = 4.2, \; \alpha_2 = 4.3, \; \alpha_3 = 4.5, \; \alpha_4 = 5.0.$}
\end{figure}
\begin{figure}[!ht]
  \centering
	
    \includegraphics[width=1.0\linewidth,height=10cm]{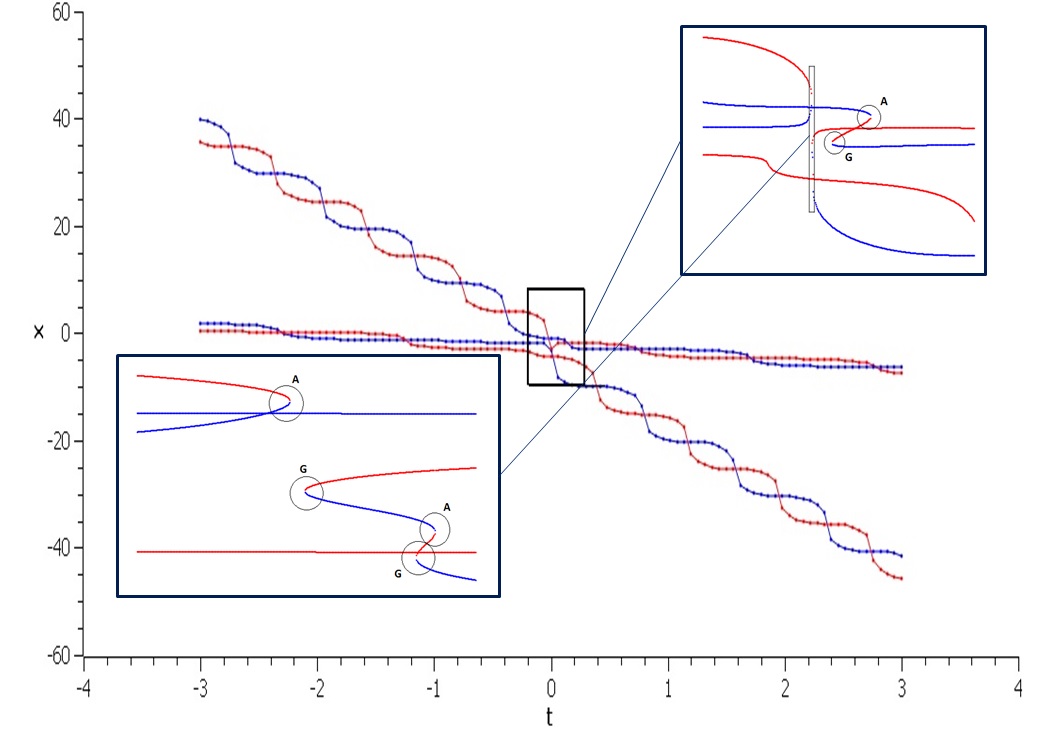}
		\label{fig:N4complexstrong}
  \caption{Singularity lines, SHG equation, N = 4, all particles of the same type; parameters: $\omega_1 = 1.0 + \imath 0.83, \; \omega_2 = 1.0 - \imath 0.83, \; \omega_3 = 3.0 + \imath 2.5, \; \omega_4 = 3.0 - \imath 2.5, \; \alpha_1 = 0.5, \; \alpha_2 = 0.7, \; \alpha_3 = 0.8, \; \alpha_4 = 0.9.$}
\end{figure}
\end{document}